\pdfoutput=1
\documentclass[traditabstract]{aa}

\usepackage{graphicx}
\usepackage{rotating}
\usepackage{txfonts}

\begin{document}

\title{A near-infrared variability campaign of TMR-1: New light on the nature of the candidate protoplanet TMR-1C}

\titlerunning{The nature of TMR-1C}

\author{B. Riaz,\inst{1} E. L. Mart\'{i}n,\inst{2} M. G. Petr-Gotzens,\inst{3} J.-L. Monin\inst{4} }
\institute{Centre for Astrophysics Research, Science \& Technology Research Institute, University of Hertfordshire, Hatfield, AL10 9AB, UK\\
\and Centro de Astrobiolog\'{i}a (CSIC/INTA), 28850 Torrej\'{o}n de Ardoz, Madrid, Spain \\ 
\and European Southern Observatory, Karl-Schwarzschild-Str. 2, 85748 Garching bei M\"{u}nchen, Germany \\
\and UJF-Grenoble/CNRS-INSU, Institut de Plan\'{e}tologie et d'Astrophysique de Grenoble (IPAG) UMR 5274, 38041, Grenoble, France \\
}

\date{Received --- ; Accepted ---}

\abstract{We present a near-infrared (NIR) photometric variability study of the TMR-1 system, a Class I protobinary located in the Taurus molecular cloud. Our aim is to confirm NIR variability for the candidate protoplanet, TMR-1C, located at a separation of about 10$\arcsec$ ($\sim$1000 AU) from the protobinary. We conducted a photometric monitoring campaign between October, 2011, and January, 2012, using the CFHT/WIRCam imager. We were able to obtain 44 epochs of observations in each of the $H$ and $K_{s}$ filters, resulting in high quality photometry with uncertainties of less than one-tenth of a magnitude. The shortest time difference between two epochs is $\sim$14 minutes, and the longest is $\sim$4 months. Based on the final accuracy of our observations, we do not find any strong evidence of short-term NIR variability at amplitudes of $\geq$0.15--0.2 mag for TMR-1C or TMR-1AB. Our present observations, however, have reconfirmed the large-amplitude long-term variations in the NIR emission for TMR-1C, which were earlier observed between 1998 and 2002, and have also shown that no particular correlation exists between the brightness and the color changes. TMR-1C became brighter in the $H$-band by $\sim$1.8 mag between 1998 and 2002, and then fainter again by $\sim$0.7 mag between 2002 and 2011. In contrast, TMR-1C has persistently become brighter in the $K_{s}$-band in the period between 1998 and 2011. The ($H-K_{s}$) color for TMR-1C shows large variations, from a red value of 1.3$\pm$0.07 and 1.6$\pm$0.05 mag in 1998 and 2000, to a much bluer color of -0.1$\pm$0.5 mag in 2002, and then again a red color of 1.1$\pm$0.08 mag in 2011. The difference in the variability trends observed in the $H$ and $K_{s}$ bands suggests the presence of more than one origin for the observed variations. The observed variability from 1998 to 2011 suggests that TMR-1C becomes fainter when it gets redder, as expected from variable extinction, while the brightening observed in the $K_{s}$-band could be due to physical variations in the inner disk structure of TMR-1C. 

%We have argued against the scenario that TMR-1C could be a faint background star, and the large-amplitude yearly variations are observed as it passes behind some patchy foreground material lying in the line of sight. 

We have argued in favour of TMR-1C being a young stellar object (YSO), rather than a faint background star passing behind some foreground material. There may exist short-term NIR variations at an amplitude level lower than our detection limit ($\sim$ 0.2 mag), which would be consistent with the YSO hypothesis. If the observed long-term variability is due to foreground extinction, then we would expect simultaneous brightening/dimming in the $H$ and $K_{s}$ bands, which is not found to be the case. Variable foreground extinction is also expected to occur over a large spatial scale; we have monitored several other objects within 4$\arcmin$$\times$4$\arcmin$ of the TMR-1 system, and found only two objects which show long-term variations, indicating that this is not a large-scale effect. The NIR colors for TMR-1C obtained using the high precision photometry from 1998, 2000, and 2011 observations are similar to the protostars in Taurus, suggesting that it could be a faint dusty Class I source. TMR-1C is a strong candidate YSO, but final confirmation as a protoplanet remains elusive and requires further investigation.

Our study has also revealed two new variable sources in the vicinity of TMR-1AB, which show long-term variations of $\sim$1--2 mag in the NIR colors between 2002 and 2011. The proper motions measured for TMR-1AB and TMR-1C are -40,+58 mas/yr and -22,+5 mas/yr, respectively, with an uncertainty of $\sim$31 mas/yr. A larger baseline of 20 years or more is required to confidently confirm the physical association of TMR-1AB and C.

}

\keywords{Stars: individual (TMR-1, TMR-1AB, TMR-1C) -- Stars: protostars -- Stars: planetary systems -- Stars: variable -- Stars: pre-main-sequence}

\maketitle

\section{Introduction}

A notable characteristic of young pre-main sequence stars is photometric variability. Optical and near-infrared (NIR) monitoring observations can probe temperature, opacity, and geometrical variations in the circumstellar environment of a young star. Photometric variability studies conducted in several clusters have shown that young stellar objects (YSOs) exhibit both periodic and aperiodic variations, with amplitudes of 0.1-1 mag and timescales of minutes to years (e.g., Bouvier et al. 1993; Herbst et al. 1994; Kenyon \& Hartmann 1995; Carpenter et al. 2001). The periodic variations yield information on stellar rotational periods and the sizes and temperatures of cool magnetic or hot accretion spots on the stellar photosphere (e.g., Bouvier et al. 1993). Aperiodic variability may arise from e.g., certain temporally changing inhomogeneities in the circumstellar environment or the ambient molecular cloud, or due to an irregular mass accretion rate (e.g., Herbst et al. 1994; Carpenter et al. 2001). 

%We present results from a NIR variability campaign of the TMR-1 system. 

TMR-1 (IRAS 04361+2547) is a deeply embedded Class I protostellar system located in the Taurus molecular cloud. Near-infrared imaging and millimeter interferometry show a bipolar outflow extending southeast to the northwest (Terebey et al. 1990). This protostar has an estimated stellar mass of $\sim$0.5$M_{\sun}$, a bolometric luminosity of $\sim$3$L_{\sun}$, and an $A_{V}$ of 28$\pm$2 mag (Terebey et al. 1990). Terebey et al. (1998; hereafter T98) presented {\it HST}/NICMOS observations of TMR-1, and were able to resolve this protostar into two point sources, A and B (the northern component was named A), at a separation of 0.31$\arcsec$. A third fainter point source, TMR-1C, was detected at a separation of about 10$\arcsec$ ($\sim$1000 AU) from the protobinary. Assuming that TMR-1C is located in Taurus and is at the same age as TMR-1AB, T98 estimated a bolometric luminosity of approximately 10$^{-3}$$L_{\sun}$ and a mass of 2-5 $M_{J}$ for this source, thus classifying it as a candidate protoplanet. Their observations also revealed a narrow filament-like structure extending southeast of the central proto-binary system towards TMR-1C, based on which T98 suggested a morphology in which the candidate protoplanet may have been ejected from the TMR-1 system, with the filament traversing the ejection path. However, follow-up low-resolution ({\it R} $\sim$120) NIR spectroscopy obtained by Terebey et al. (2000) with the Keck/NIRC instrument showed a featureless spectrum, and it was suggested that this may be a background star. 

Two recently published studies by Riaz \& Mart\'{i}n (2011) and Petr-Gotzens et al. (2010) strongly argued that TMR-1C is most likely a YSO, possibly surrounded by a highly inclined disk, and not an extincted background star. In Riaz \& Mart\'{i}n (2011), we reported large-amplitude variations of $\sim$1--2 mag for TMR-1C in the $H$- and $K_{s}$-bands, over a $\sim$4 year time period. We also noted a reddening in the ($H-K_{s}$) color as the object gets fainter in both bands. However, our previous results were based on a single epoch of NIR observations, and the large photometric uncertainties made it difficult to reliably confirm TMR-1C as a variable source. 

%We conducted a $\sim$4 month long NIR imaging campaign of the TMR-1 system, and 

In our present NIR imaging campaign of the TMR-1 system conducted between October, 2011, and January, 2012, we have obtained 44 epochs of observations in each of the $H$ and $K_{s}$ filters, resulting in higher quality photometry with uncertainties a factor of $\sim$10 smaller than previous measurements. Our present campaign covers shorter timescales of $\sim$10 minutes to $\sim$4 months, making it possible to probe short-term variability for the TMR-1 system. Also, a comparison with the previous data sets provides a $\sim$9 year baseline to measure the proper motion and determine Taurus membership for the TMR-1 components. Section \S\ref{obs} provides details of the observing campaign, an analysis of the photometric and astrometric data is described in Section \S\ref{analysis}, results on short- and long-term NIR variability are given in Section \S\ref{results}, while Section \S\ref{origins} discusses the possible origins of variability in TMR-1C. 

%Near-infrared (NIR) monitoring observations can probe temperature, opacity, and geometry changes in the circumstellar environment. Photometric variability studies conducted in several clusters have shown that young stellar objects (YSOs) exhibit both periodic and aperiodic variations, with amplitudes of 0.1-1 mag (e.g., Bouvier et al. 1993; Herbst et al. 1994; Kenyon \& Hartmann 1995; Carpenter et al. 2001). 

%Characteristics of protoplanets

%how are wide binaries formed? Formation mechanism... Fraction of such sources with a planetary mass companion. 

\section{Observations}
\label{obs}

Our NIR imaging campaign was conducted between October, 2011, and January, 2012. Observations were obtained at the CFHT 3.6m telescope, using the wide-field infrared camera, WIRCam. This imager covers a 20$\arcmin\times20\arcmin$ field of view, with a pixel scale of 0.3$\arcsec$/pix. We obtained alternate $H$ and $K_{s}$ observations, so as to check for short-term variability in each filter. In both bands, a 5-point dithering pattern (20$\arcsec$ dithering offset) was used, with 5 exposures obtained at each dither position. The exposure time was set to 15 sec in the $H$-band, and 20 sec in the $K_{s}$-band. Combining the various overheads, a $\sim$14 minute observing time was required to complete one set of observations (total 25 frames) in each filter. Thus, we were able to obtain 4 epochs of observations per filter in 2 continuous hours of observing time during one night. This sequence allowed us to probe short-term variability over a few minutes to an hour timescale. In order to probe variability over a week-long and a monthly period, the 2-hour sequence was repeated every night for 7 consecutive nights during October, 2011, and then 4 times during the 2011A semester. In this way, we were able to obtain a total of 88 epochs, i.e., 44 epochs per filter, over the full observing campaign. The shortest time difference between two sets of observations in a single filter is $\sim$14 minutes, and the longest is $\sim$4 months. In January 2012, 4 nights were affected by a partial cloud cover, with poor seeing of 0.85$\arcsec$-1.1$\arcsec$. The rest of the nights were photometric, with stable seeing between 0.5$\arcsec$ and 0.8$\arcsec$. 

The raw images were pre-processed by the WIRCam observer team using the `I'iwi data reduction pipeline (version 2.1.1). The `detrended' images obtained from the pipeline first have all detector imprints removed, followed by dark subtraction, flat-fielding, and sky-subtraction. The crosstalk is removed from the final pre-processed images, and astrometry performed with an rms scatter of the resulting WCS solution between 0.3$\arcsec$ and 0.8$\arcsec$. To obtain a good signal-to-noise ratio (SNR), we stacked the 25 images taken over $\sim$10 minutes in one set of observations, for each filter. TMR-1C is clearly detected in the stacked images in both bands. For the protostar TMR-1AB, we could not resolve it into its individual components, and all photometric measurements are for the composite source. The images shown in Figs.~\ref{spatial-full} and \ref{spatial-AB} are from a deep stack constructed by combining all 44 stacks. 

%In total, we have 56 stacks in the $H$-band, and 54 in the $K_{s}$-band. 

\section{Data Analysis}
\label{analysis}

\subsection{Differential Photometry}
\label{diffM}

\begin{figure*}    
\centering
      \includegraphics[width=150mm,height=130mm]{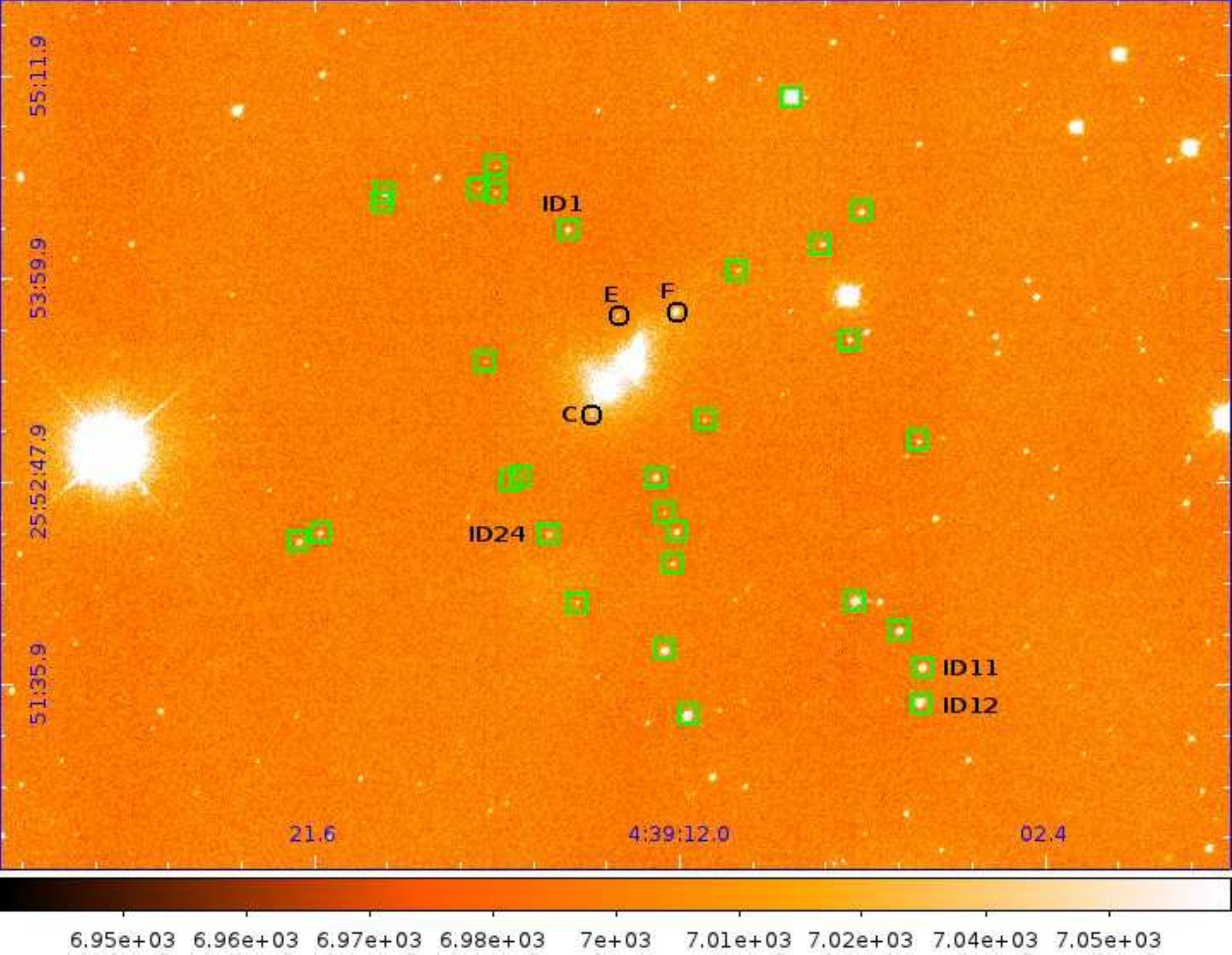}           
     \caption{The WIRCam 2011 $H$-band false color image (stack of all 44 epochs), with the reference star sample marked as open boxes. Also denoted are some sources which show long-term variability. North is up, east is to the left. The scale bar at the bottom is in units of MJy sr$^{-1}$. } 
    \label{spatial-full}
 \end{figure*}

The first step in our analysis was to conduct differential photometry, and compare the rms values for the targets (TMR-1 AB and C) with the mean rms derived for a reference star sample. We selected 30 reference stars in an area of 4$\arcmin \times$4$\arcmin$ roughly centered at TMR-1AB (Fig.~\ref{spatial-full}). Aperture photometry was performed for the reference and target stars in each of the stacked images using the photometry tasks under the IRAF {\em digiphot} package. We chose 4 different aperture radii between 1 and 10 pixels, and a sky annulus with an inner radius of 3 pixels larger than the aperture radius, and a width of 4 pixels. For the components AB and C, as well as the two other objects that lie close to the nebulous region around AB (marked as `E' and `F' in Fig.~\ref{spatial-full}), we used a different method for sky subtraction, as discussed further in \S\ref{phot-AB}. The photometry was calibrated using 2MASS magnitudes for bright sources in the same field, for every epoch. The radial profiles for these bright sources were checked for any saturation effects. For each reference star, a differential magnitude was calculated by subtracting from the calibrated magnitude the weighted mean magnitude of all of the other reference stars, in each stacked image. We did various checks on the stability of the mean magnitude of the reference star sample. In each epoch, we have 30 weighted mean magnitudes. The scatter in those mean magnitudes is between $\sim$0.05 and 0.065 mag. We were able to reduce this scatter to less than 0.05 mag by deleting one particular epoch (or one particular stack) which has bad image quality (high extinction as noted in the observing log). We also deleted the reference stars ID1, ID11, and ID24, which may be marginally variable in the $H$ or the $K_{s}$-band (Section \S\ref{variability}). We note that deleting these reference stars does not significantly reduce the rms, and the difference is found to be of $\sim$0.004 mag in the resulting standard deviation of the mean magnitude. Another way to reduce the scatter and obtain a stable reference sample is to delete the 8 faintest stars, all of which have calibrated magnitudes of $>$18 mag, and the photometry errors are also larger, $>$0.07 mag. However, rejecting these sources will significantly reduce the size of a comparative sample at these faint magnitudes, which we require in order to check for the variability of C. These faint reference stars were thus not excluded form the sample. We have applied the zero-point offset separately for each epoch, which should also result in a stable reference sample. 

\begin{figure*}    
 \centering
      \includegraphics[width=80mm]{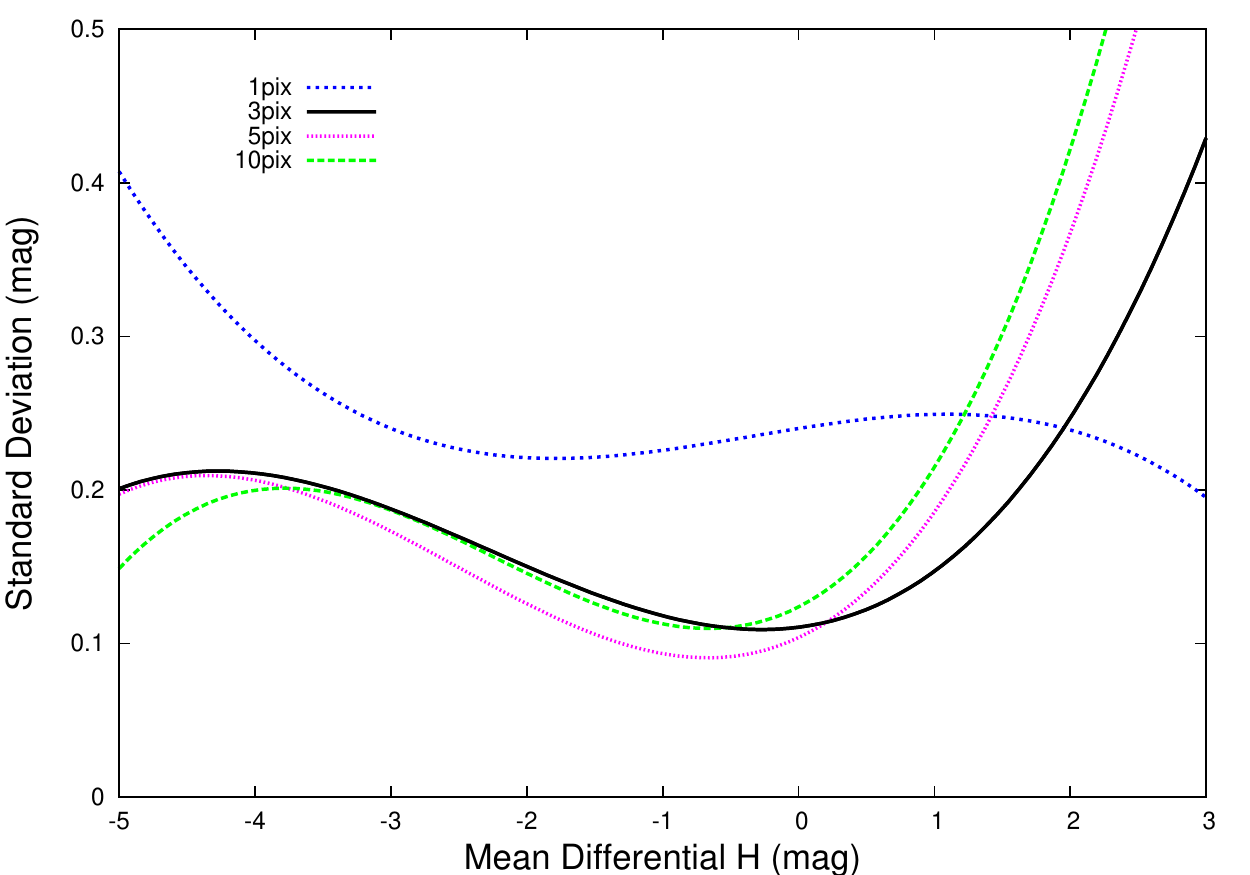}           
       \includegraphics[width=80mm]{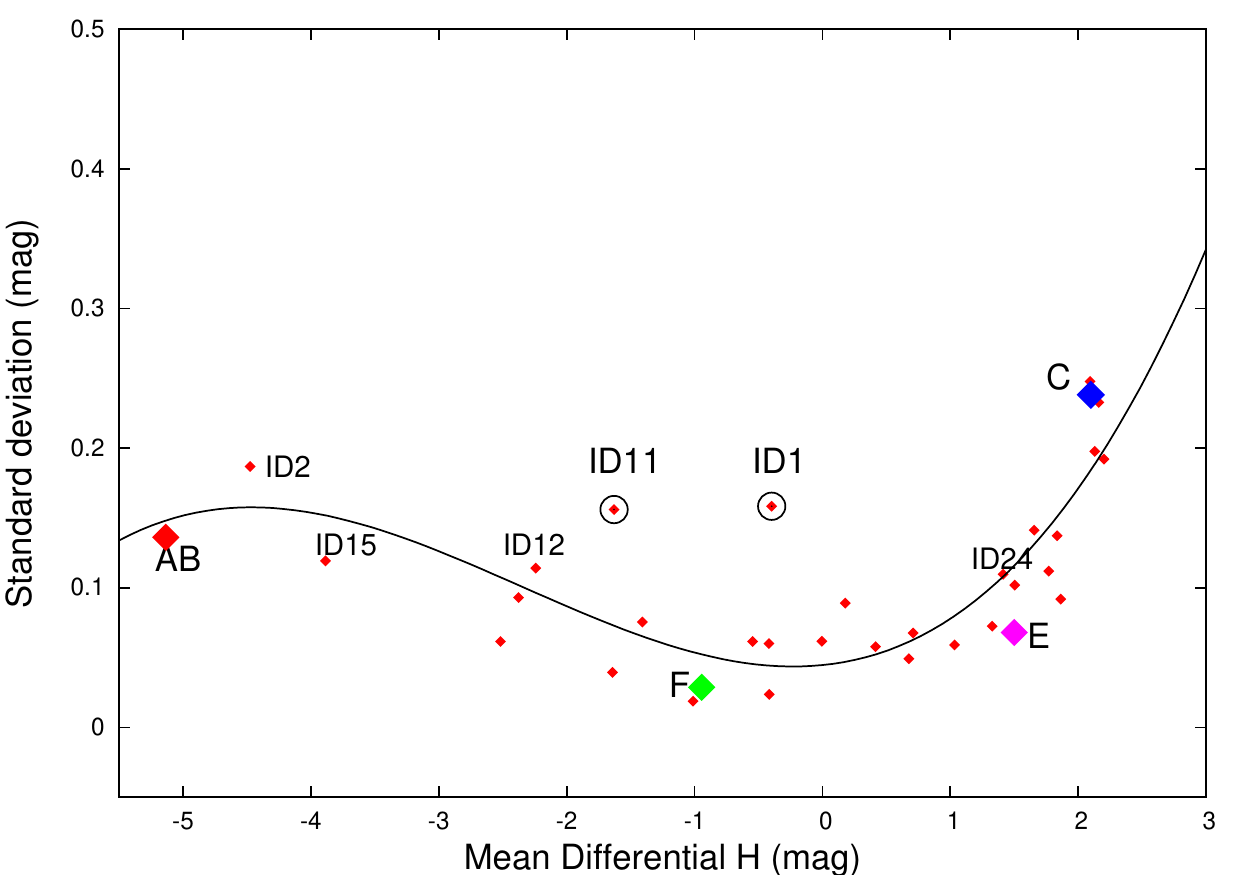}         
      \includegraphics[width=80mm]{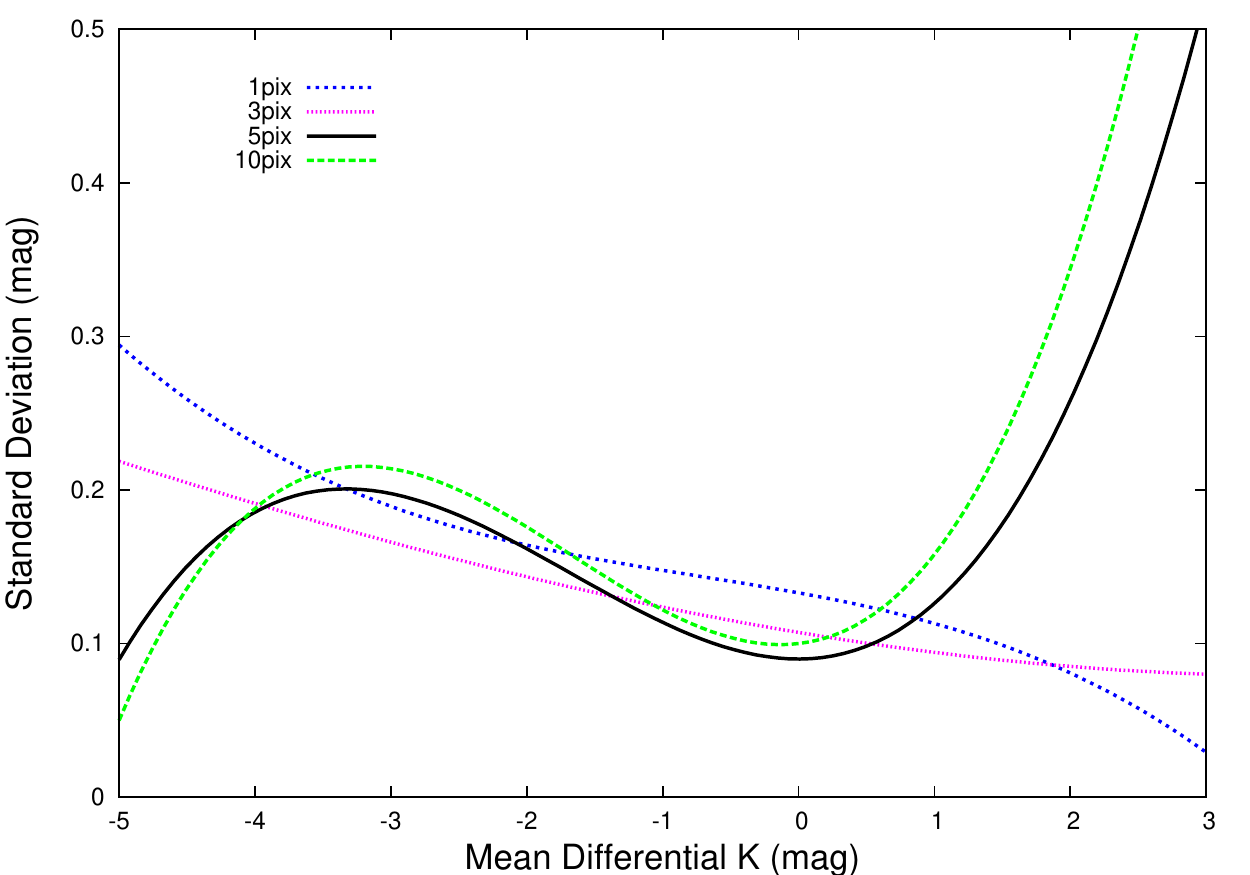}     
      \includegraphics[width=80mm]{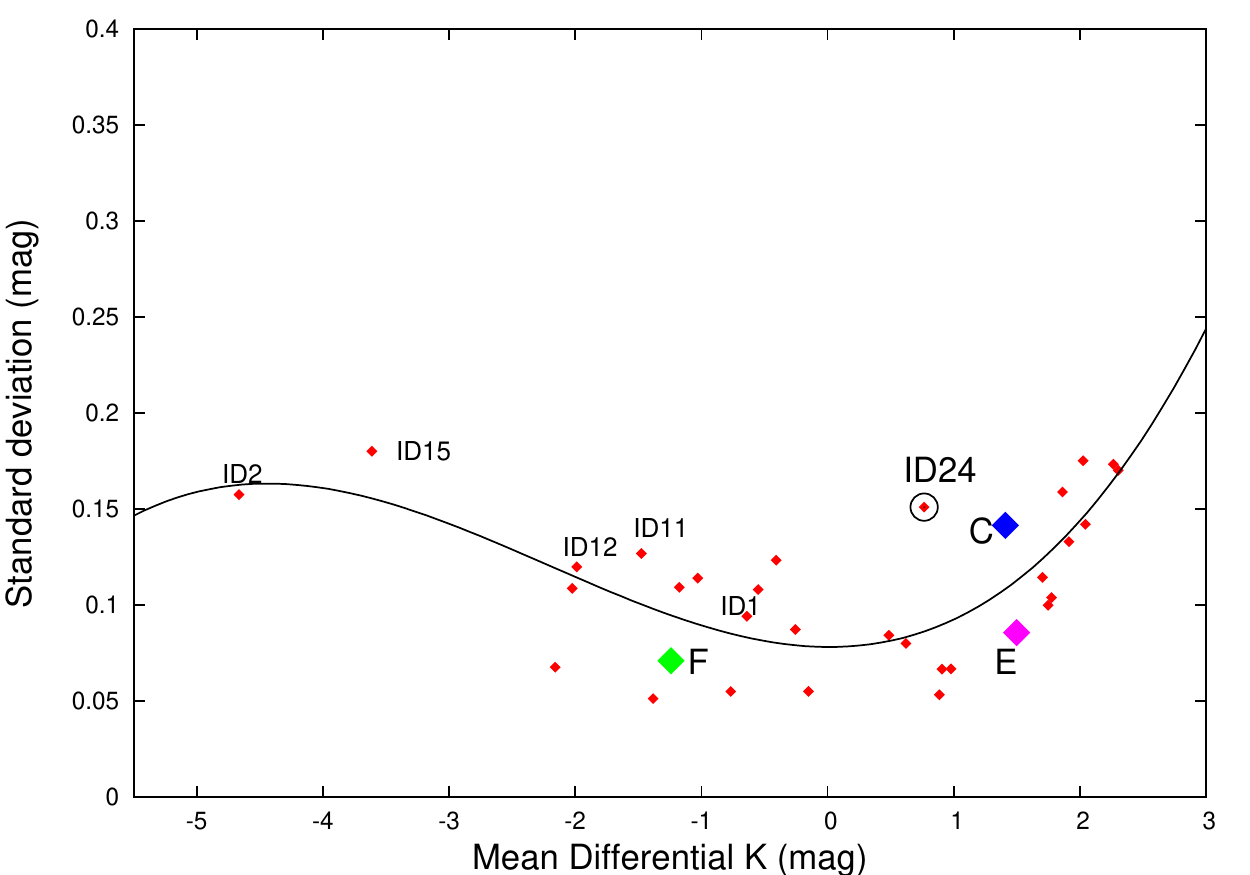}         
     \caption{{\it Left panel}: (a) The fitting function to the standard deviation versus the mean differential magnitude for the reference star sample, using aperture radius of 1, 3, 5, and 10 pixel. Top panel shows the fits for the $H$-band, bottom panel for the $K_{s}$-band. Black line represents the aperture radius considered for the photometry in the two bands. {\it Right panel}: (b) The standard deviation versus the mean differential magnitude for the reference star sample and the targets, in the $H$-band (top) and the $K_{s}$-band (bottom). Candidate variables are marked with black circles.  }  
    \label{apfits}
 \end{figure*}

We then calculated the mean and the standard deviation of the distribution of differential magnitudes for each reference star from all stacked images. For AB and C, the differential magnitudes were calculated by subtracting the mean calibrated magnitude for all 30 stars in the reference sample. This method was repeated for all 4 aperture radii. We then plotted the mean differential magnitude versus the standard deviation for the reference star sample, and used a polynomial function to fit the distribution of points. Figure~\ref{apfits}a shows the fits in the $H$- and $K_{s}$-band for the different aperture radii used. Our reference sample consists of stars with a range in magnitude between $\sim$12 and 18 mag in the $H$-band, and between $\sim$11.3 and 18.2 mag in the $K_{s}$-band. The sigma values and thus the fits in Fig.~\ref{apfits}a are expected to rise towards fainter objects, i.e. for more positive values for the differential magnitude. We see this rise for aperture radii of 3 pixel or larger in the $H$-band, and 5 pixel or larger in the $K_{s}$-band, whereas for smaller apertures, the fits are nearly flat. The 3 pixel aperture gives the best compromise across the full range of magnitudes in the $H$-band, whereas this is found to be the case for the 5 pixel aperture in the $K_{s}$-band. Therefore, the final photometry considered for the reference sample and the targets is that obtained using a 3 pix and a 5 pix aperture radius in the $H$- and $K_{s}$-band, respectively. The appropriate aperture correction was determined from the curve of growth of the brightest star in the reference sample, and added to the calibrated photometry. The presence or absence of variability based on the differential photometry plots is discussed in Section \S\ref{variability}.

%These aperture radii are comparable to the average FWHM (2.8--3.3) of the reference stars in the respective bands. 

\subsection{Photometry for TMR-1AB and C}
\label{phot-AB}

\begin{figure*}    
\centering
      \includegraphics[width=100mm]{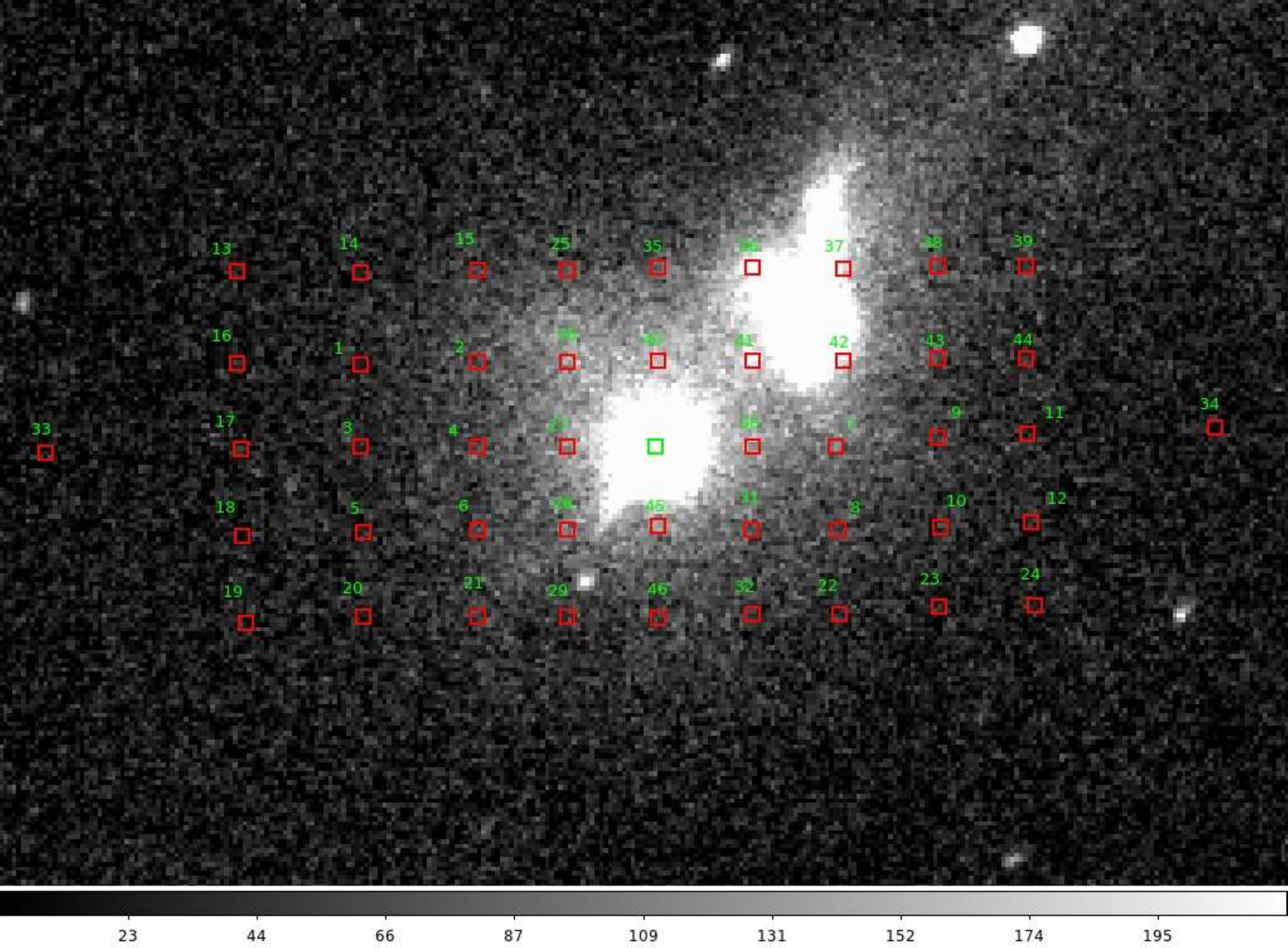} \\              
      \includegraphics[width=80mm]{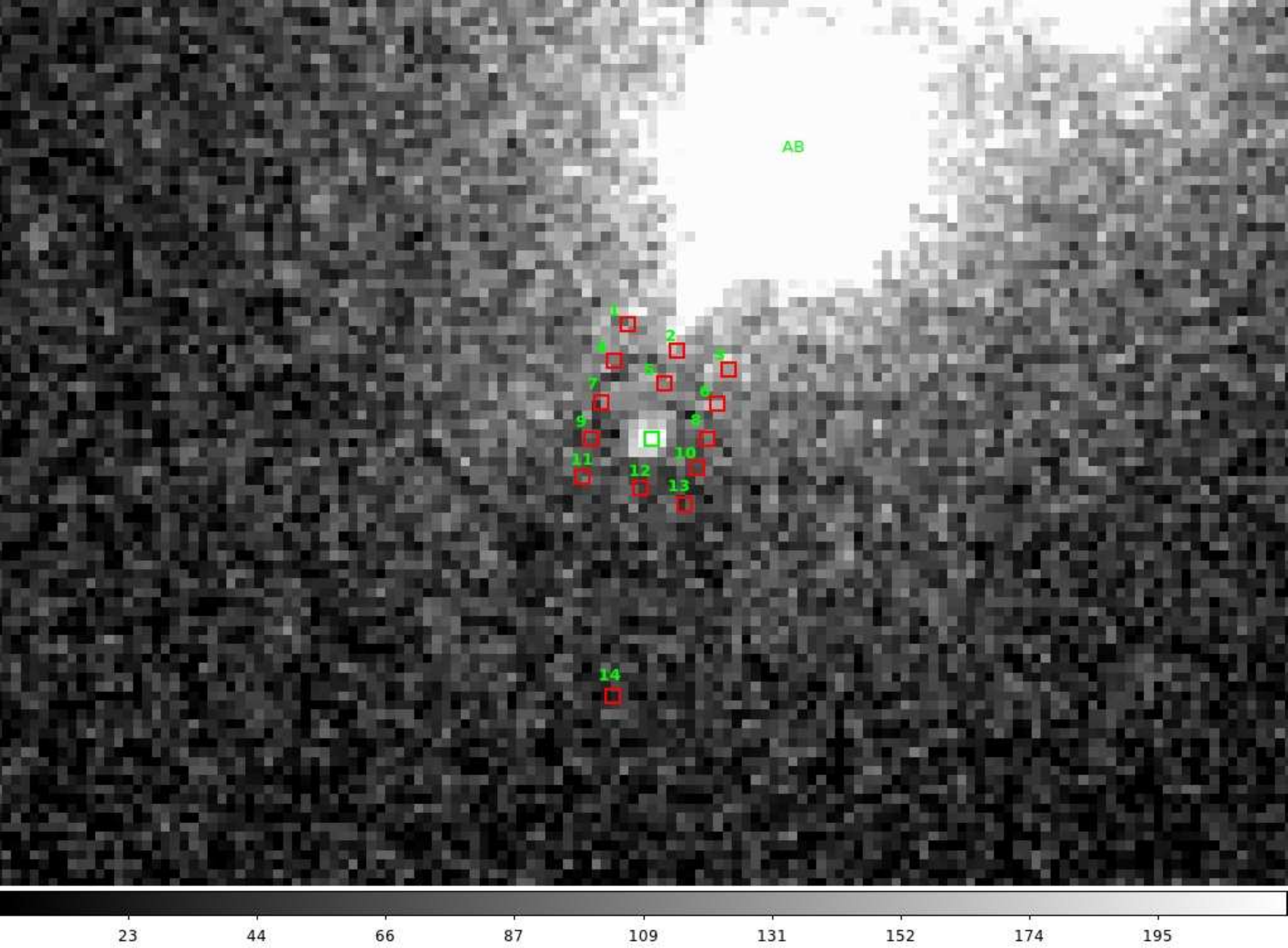} 
      \includegraphics[width=80mm]{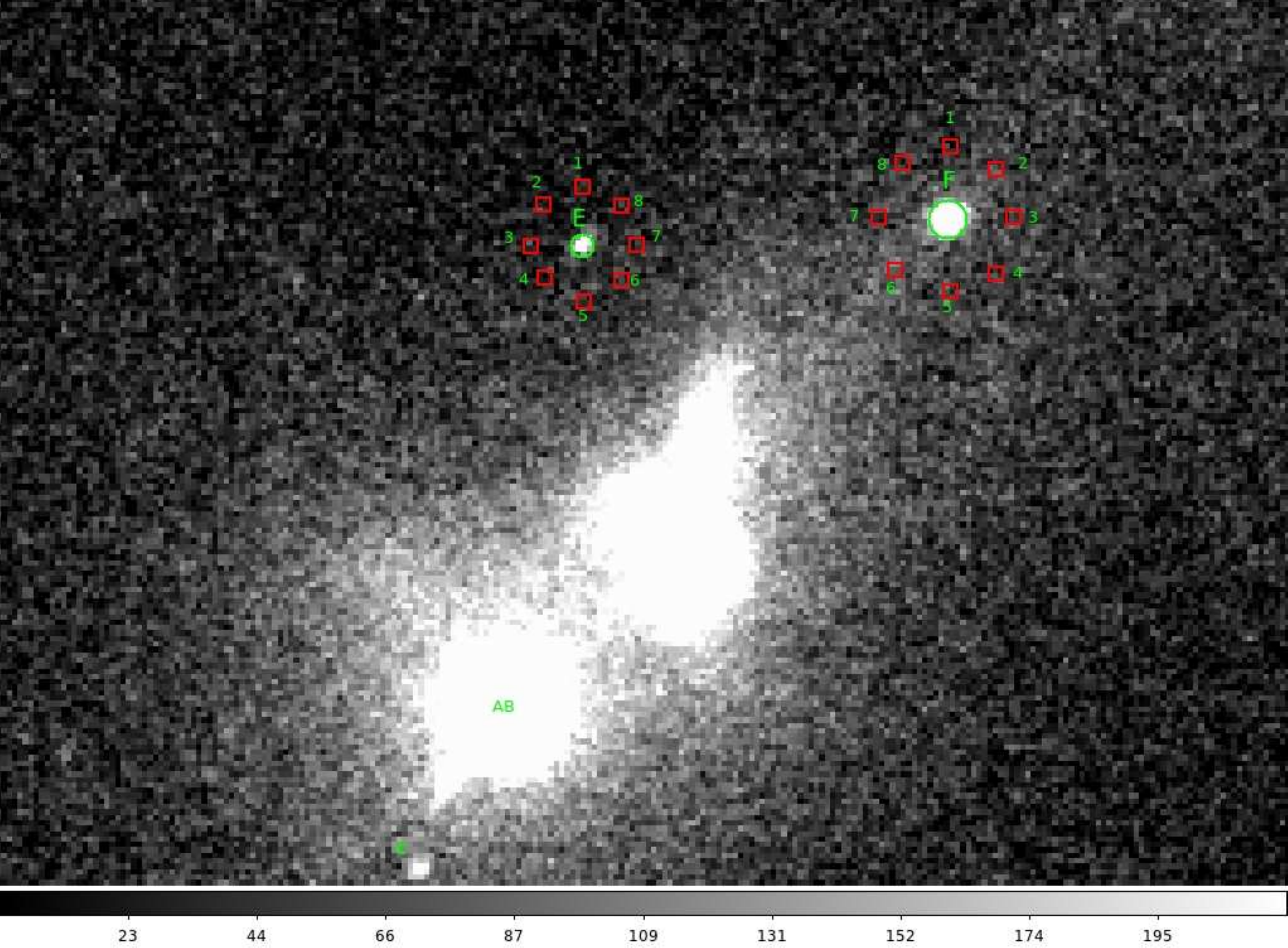}             
     \caption{{\it Top}: (a) The grid of sky locations (marked by red squares) selected both inside and outside the nebulous region to measure the photometry for TMR-1AB (marked by a green square). {\it Bottom, left}: (b) The sky locations selected close to and farther away from the nebulous region surrounding AB, to measure the photometry for C (green square). {\it Bottom, right}: (c) The sky locations used to measure the photometry for E and F, which lie north of TMR-1AB. These image cuts are from the stack of all 44 epochs of WIRCam 2011 $H$-band images; north is up, east is to the left. The scale bar at the bottom of each image is in units of MJy sr$^{-1}$. } 
    \label{spatial-AB}
 \end{figure*}

The photometry for the components AB and C can be expected to have some contamination from the surrounding nebulosity. We therefore employed a different method for sky subtraction for these components, in addition to using a typical sky annulus of a certain width. We selected various `sky locations' around these objects, and then measured the photometry using the sky value from each of the sky locations (Fig.~\ref{spatial-AB}, top panel). The sky background was measured using the same aperture radius as used for the target objects. The various magnitudes obtained using this method were then compared to the photometric measurement obtained at the location of the target, using a sky annulus of 3 pixels larger than the aperture radius and a width of 4 pixels. For AB, a grid of sky locations was selected such that there were points both inside and outside the nebulous region (Fig.~\ref{spatial-AB}). The photometry for AB is slightly brighter when subtracting sky values from a less nebulous region, compared to the magnitude at the source location. The full range in magnitudes for AB from all sky locations is only $\sim$0.02 mag, which suggests that the surrounding nebulosity is evenly and relatively smoothly distributed around the source at a few pixel level. We can roughly estimate the nebulosity surface brightness per pixel by this full range in magnitude divided by the area in pixels considered for the measurements. This is found to be $\sim$0.0008 mag/pixel, which is a negligible value. The main contribution to the photometry for AB is thus from the protostar itself, and the surrounding nebulosity does not prevent a precise measurement of the photometry. We therefore considered the photometric measurement for AB obtained at the source location, using a sky annulus of 30-40 pixels. We note that AB appears extended and shows a saturated radial profile for all single images in the $K_{s}$-band, due to which we do not have a photometric measurement in this band. The final calibrated and aperture corrected $H$-band photometry for AB, obtained by averaging the photometry from all 44 epochs, is listed in Table~\ref{photC}. 

%, including the ones which are at a distance of $>$30 pixels from the source position

%The photometry for AB is also found to be constant for sky locations which are at a distance of $>$30 pixels from the source position. In other words, for any point outside the nebulous region, the resulting magnitudes are the same, whereas the magnitudes are brighter by $\sim$0.02 mag when using sky measurements at locations inside the nebulous area. This difference of $\sim$0.02 mag is very small, and is comparable to the uncertainty on the photometry. 

% similar results because the nebulosity is spatially smooth. you could have a reasonably bright nebulosity with a source superimposed on it, the annulus method to subtract the source surrounding emission would give the same result, provided its not to bright and adds statistical "signal" noise. 

%The photometry for the components AB and C, and the objects E and F which lie close to the nebulous region around AB (Fig.~\ref{spatial2}), can be expected to have some contamination from the surrounding nebulosity. 

%The large sky annulus takes into account the nebulosity. 

The contamination is comparatively more enhanced for the much fainter target TMR-1C. For the component C, we selected various sky locations that lie close to and farther away from the nebulous region surrounding AB, as marked in Fig.~\ref{spatial-AB} (bottom left panel). Since the nebulous emission is stronger at locations 1--6, the magnitudes for C obtained using sky background measurements at these locations is $\sim$1.7 mag fainter than the magnitude obtained at the source location. On the other hand, at sky locations surrounding C (locations 7--10), or the ones located farther away from the nebulous region (locations 11--14), the magnitudes are brighter by $\sim$0.4 mag, compared to the magnitude at the source location. We can then estimate that the contamination due to the nebulous region at the location of C and its immediate surroundings is about 0.4 mag. To correct for the possible contamination effects, we took the mean of the sky measurements from the locations of 7--13, and then used this sky value to measure the photometry for C. We note that if we average all of the sky locations, the resulting magnitude is $\sim$0.06 mag fainter than averaging the 7-13 locations, which is a difference comparable to the photometric uncertainty. The difference between the two measurements is small because the low sky background at locations south or close to C cancel the effect of the comparatively higher sky values close to the nebulous region. Table~\ref{photC} lists the final calibrated and aperture corrected photometry for C, obtained by averaging the photometry from all 44 epochs, in both bands.

%The final magnitude which we have considered for C is the mean of the photometry obtained using the sky measurements at the neighboring locations of 7--13. 

%We can roughly estimate the nebulosity surface brightness per pixel by this full range in magnitude divided by the area in pixels considered for the measurements. This is found to be $\sim$0.0008/pixel, which is a negligible  value. 

%The photometry obtained using sky measurements at these locations was then compared with the photometry 

%The various magnitudes obtained using this method were then compared to the photometric measurement obtained at the location of the target, using a sky annulus of 3 pixels larger than the aperture radius and a width of 4 pixels. 

We also redid the photometry for AB and C from the 2002 and 2009 CFHT observations, presented in Riaz \& Mart\'{i}n (2011), using the same analysis methods as employed for the 2011 data. The new photometric measurements are comparable within the uncertainties with the ones reported in our previous work. The uncertainties are a factor of $\sim$10 smaller for the 2011 data compared to 2002 or 2009, which can be expected considering that the present measurements have been obtained from nearly 1000 images in each filter. The present photometry for TMR-1AB and C is thus of a much higher quality and is a vast improvement over all previously reported photometric measurements for these sources, including the T98 {\it HST} observations. 

Persson et al. (1998) developed a grid of {\it J-, H-, K-}, and $K_{s}$-band standards for the {\it HST} NICMOS camera, using observations from the Las Campanas Observatory (LCO) in Chile. Later, Carpenter (2001) derived 2MASS-LCO transformation equations using 2MASS photometry for 82 stars from Persson et al. (1998). Equations (1) and (2) below are the color-transformation relations from Persson et al. (1998) and Carpenter (2001), respectively. 

\begin{equation}
(H-K)_{CIT} = (0.974 \pm 0.020)(H-K)_{LCO} \pm (0.013) 
\end{equation}

\begin{equation}
(H-K_{s})_{2MASS} = (1.019 \pm 0.010)(H-K)_{LCO} + (0.005 \pm 0.005)
\end{equation}

In Tables \ref{photC} and \ref{photC}, we have listed our CFHTIR and WIRCam 2MASS-calibrated ($H-K_{s}$) color-transformed to $(H-K_{s})_{LCO}$, using Eq. (1). The results are similar using Eq. (2). We find a difference in the $(H-K_{s})$ color of $\sim$0.01 mag for TMR-1C, and 0.02-0.03 mag for TMR-1AB.

%Table~\ref{photC} lists the photometry from the 2002, 2009, and the present 2011 observations, along with the photometry reported by T98 from their {\it HST}/NICMOS observations, and the 2MASS photometry. 

\subsection{Objects in the vicinity of TMR-1AB}

In Fig.~\ref{spatial-AB} (bottom, right panel), we have marked two objects, E and F, which lie close to the nebulous region around AB. The coordinates for these objects are listed in Table~\ref{astroEF}. The object F is a 2MASS source with the designation 2MASS J04391199+2553490. There is no spectral classification available for this object. It shows significant variability in the ($H-K_{s}$) color when comparing the 2MASS and our WIRCam or CFHTIR photometry (Table~\ref{photEF}), as discussed further in Section \S\ref{longvar}. The object E has no matches in 2MASS or in SIMBAD. This is a new detection, and as discussed in Section \S\ref{astrometry}, it is a strong candidate for Taurus membership based on its proper motion. For both of these objects, we used the same photometry method as employed for AB and C. The sky locations are shown in Fig.~\ref{spatial-AB}. For E, there is a $\sim$0.3 mag difference between the photometry at the source location, and that obtained using sky measurements at locations 5--8, i.e. closer to the nebulous region. For F, there is a much smaller difference of $\sim$0.06 mag between the magnitude at the source location and that obtained using sky measurement at any of the 1--8 locations. The final photometry for these two sources was obtained using the mean sky value from all sky locations (Table~\ref{photEF}). 

%, and it has a 2MASS classification of an `infrared source'

%A similar photometry method was used for the photometry of the objects E and F, which lie close to the nebulous region around AB (Fig.~\ref{spatial2}). 

We cannot confirm the detection at a $>$2-$\sigma$ level for the faint object TMR-1D, reported in Petr-Gotzens et al. (2010). This source lies to the northwest of TMR-1AB, near the edge of a filamentary structure extending northwest from the protostar. TMR-1D is $\sim$1 mag fainter than TMR-1C in the $H$ and $K_{s}$ bands (Petr-Gotzens et al. 2010). It is undetected in our 2002 and 2009 observations. In the present data, a faint blob is seen in 6 $K_{s}$-band images from January 2012 at the TMR-1D location marked in Petr-Gotzens et al. (2010). However, the detection significance is very low and the radial profile of the blob shows only scattered emission. Deeper observations are required to conduct photometry and astrometry for this source.

%D, G detected in ... cannot confirm the detection of D at a $>$2-$\sigma$ level. In 6 images, a faint extended blob is seen at the location marked as D by Petr...no astrometry or photometry given (except it's 1.5 mag fainter than C). 

\subsection{Astrometry}
\label{astrometry}

\begin{figure*}  
\centering
      \includegraphics[width=140mm]{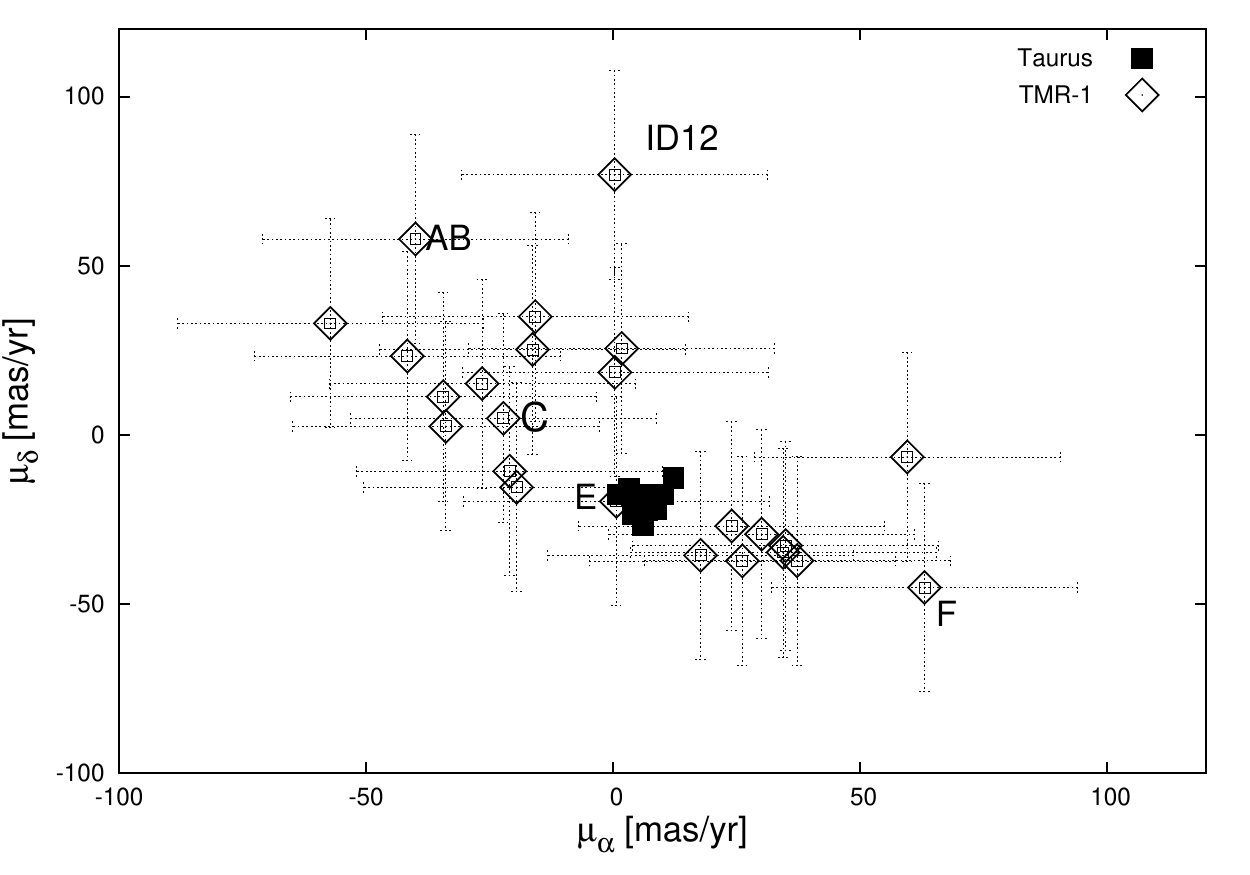}           
     \caption{Proper motion in Dec versus the proper motion in RA. Proper motions were obtained from the 2002 and 2011 $H$-band positions. Black squares denote the median proper motions for the 11 groups in Taurus from Luhman et al. (2009). The errors on proper motions for these groups are between 1 and 2 mas/yr.   } 
    \label{pm}
 \end{figure*}

The 2002 and 2011 data provide a $\sim$9 year baseline over which the proper motions for the targets can be measured. Figure~\ref{pm} plots the proper motion in RA and Dec for the reference star sample and the targets, as determined using the $H$-band observations from 2002 and 2011. Since AB is saturated in the $K_{s}$-band, we preferred to use the $H$-band data for astrometric measurements. The uncertainty of $\sim$30 mas/yr on the proper motions was calculated from the standard deviation of the full distribution of proper motions for the reference star sample. Also included for comparison in Fig.~\ref{pm} are the median proper motions for the 11 groups in the Taurus-Auriga star-forming region discussed in Luhman \& Mamajek (2009). The errors on proper motions for these groups are between 1 and 2 mas/yr. The typical proper motion for Taurus from Ducourant et al. (2005) is +2,-22 mas/yr, with uncertainties of 2-5 mas/yr.

%The proper motion for C and the reference sample obtained using the $K_{s}$ data are consistent with the $H$-band measurements. 

%The large uncertainty of $\sim$30 mas/yr on the proper motions from our 2002/2011 measurements makes it difficult to confirm the Taurus membership for the targets and the reference sample, or to confirm that AB and C are co-moving sources. 

An interesting case is the object E, which has a proper motion of +0.66,-19.6 mas/yr and overlaps with the Taurus loci (Fig.~\ref{pm}), making it a strong candidate for Taurus membership. Based on its faint magnitudes (Table~\ref{photEF}), this source is likely a brown dwarf. However, for the rest of the objects including the TMR-1 targets, the large uncertainty of $\sim$30 mas/yr on the proper motions from our 2002/2011 measurements makes it difficult to confirm Taurus membership. Fig.~\ref{pm} also shows the possible presence of two groups of stars, based on the proper motions; group A with $\mu_{\alpha} >$0 or more positive than the Taurus loci, and group B with comparatively negative $\mu_{\alpha}$ of $\leq$0, similar to AB and C. We checked if this could be a systematic effect due to the respective position of an object on the detector, but no particular correlation is observed between the XY position and the proper motion of the sources. This subgrouping may be real, but a confirmation of it would require higher precision astrometric measurements obtained over a larger baseline than the $\sim$9 years considered here. 

%suggesting that the proper motions are nearly constant across the Taurus region.

Within the large uncertainties, the proper motion for TMR-1AB and C are comparable. The protostar AB, however, shows a higher proper motion in declination compared to C (Table~\ref{astrometry}). The separation between AB and C, as measured in all of the $H$-band stacked images, is between 9.8$\arcsec$ and 11.3$\arcsec$. The mean value is 10.2$\arcsec$, with a standard deviation of 0.3$\arcsec$. This is consistent with the 10$\arcsec$ separation reported by Terebey et al. (1998), and the 9.8$\arcsec$ measurement from our 2002 observations. TMR-1AB and C likely form a common proper motion pair. With a larger baseline of 20 years or more, the astrometric error bars can be expected to be much smaller than the present estimates, and the association of AB and C can then be confidently confirmed. 

%The total proper motion for AB and C using the 2002/2011 baseline is 19.7 mas/yr and 23.6 mas/yr, respectively. 

%larger baseline... show RA vs pm with AB only -->   pm in this region consistent within error bars with other taurus loci

%the sep between AB and C are consistentely 10atcsec for all images. the pm are similar within error bars. need a larger baseline. 

\section{Results}
\label{results}

\subsection{Short-term NIR variability} 
\label{variability} 

In order to probe variability over the short-term periods of a few minutes to $\sim$4 months probed in the 2011 observations, we compared the rms values of the differential magnitudes (Section \S\ref{diffM}) for the targets with the reference star sample, in both bands. This is shown in Fig.~\ref{apfits}b. The solid line in this figure is a fit to the distribution of points for the reference stars, and represents the mean rms for this sample. We considered an object to be a variable source if the observed rms was a factor of 2 higher than the mean rms for the reference sample, at a given differential magnitude. In the $H$-band, both TMR-1AB and TMR-1C have rms values consistent with the mean for the reference sample, and thus can be considered as non-variables over this short-term period. In the $K_{s}$-band, C lies at a $\sim$1.2 sigma level above the mean rms curve, which is not significant enough to classify it as a variable. The objects E and F that lie close to the nebulous region also do not show any short-term variability in both bands.

%Fig.~\ref{apfits} shows that both TMR-1AB and TMR-1C have rms values consistent with the mean for the reference sample. These sources can therefore be considered as non-variables over the short-term periods of a few minutes to $\sim$4 months probed in the 2011 observations. The objects E and F that lie close to the nebulous region also do not show any short-term variability in both bands. 

We do find two references stars, ID1 and ID11, which lie at a $\sim$2.9-sigma and $\sim$1.9-sigma level, respectively, above the mean rms in the $H$-band differential plot (Fig.~\ref{apfits}b). The photometry for these objects is listed in Table~\ref{photEF}. We note that the time series plots for these two sources are similar to the other non-variable sources in the reference sample. None of these sources lie at a $\geq$2-$\sigma$ level above the mean rms in the $K_{s}$-band differential plot, and can be considered as non-variable in this band. ID11 shows clear signs of long-term variability over the 2002-2011 period, as discussed in Section \S\ref{newvar}, whereas ID1 is non-variable over this long-term period. 

A well-known method of identifying variable objects is based on the Stetson variability index (Stetson 1996), which is particularly reliable when checking for correlated variations in multiband magnitudes. We calculated the Stetson variability index, {\it S}, for each star from the observed $H$ and $K_{s}$ magnitudes and associated photometric uncertainties, as described in Carpenter et al. (2001). 

\begin{equation}
S = \frac{\sum_{i=1}^{p} g_{i} ~ sgn (P_{i}) ~ \sqrt{{|P_{i}|}}} {\sum_{i=1}^{n} g_{i}}, 
\end{equation}

\noindent where {\it p} is the number of epochs of observations, {\it n} is the number of measurements used to determine the mean magnitude. The parameter $P_{i}$ is the product of the normalized residuals of the observations in the two bands,

\begin{equation}
P_{i} = \frac{n}{n-1} ~~ \frac{m_{H} - \bar{m}_{H}}{\sigma_{m_{H}}} ~~~ \frac{m_{K_{s}} - \bar{m}_{K_{s}}}{\sigma_{m_{K_{s}}}},
\end{equation}

\noindent and $g_{i}$ is the weight assigned to each normalized residual. We assigned a weight of 1/2 in each band, implying a total weight of 1.0, for detections in both bands.

\begin{figure*}    
 \centering
      \includegraphics[width=110mm]{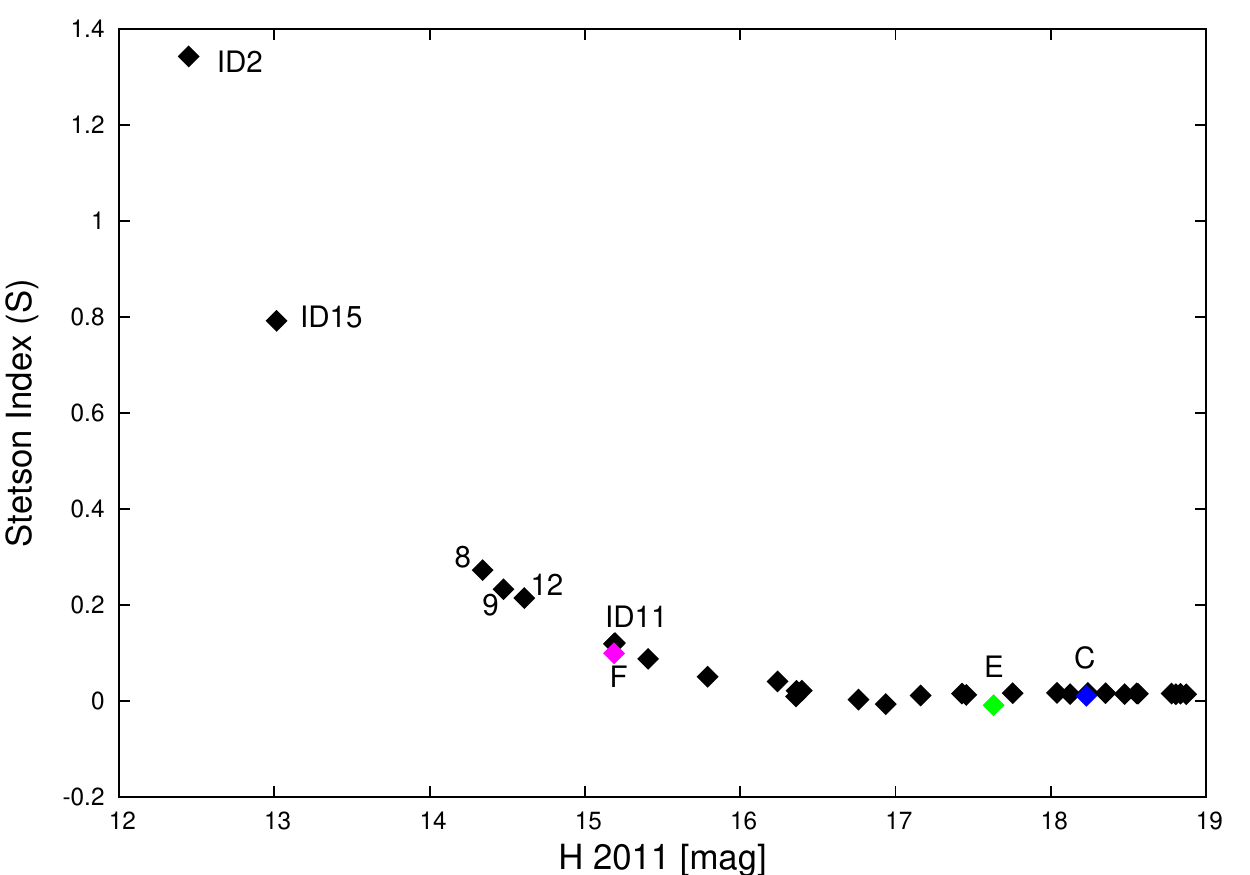} \\
      \includegraphics[width=80mm]{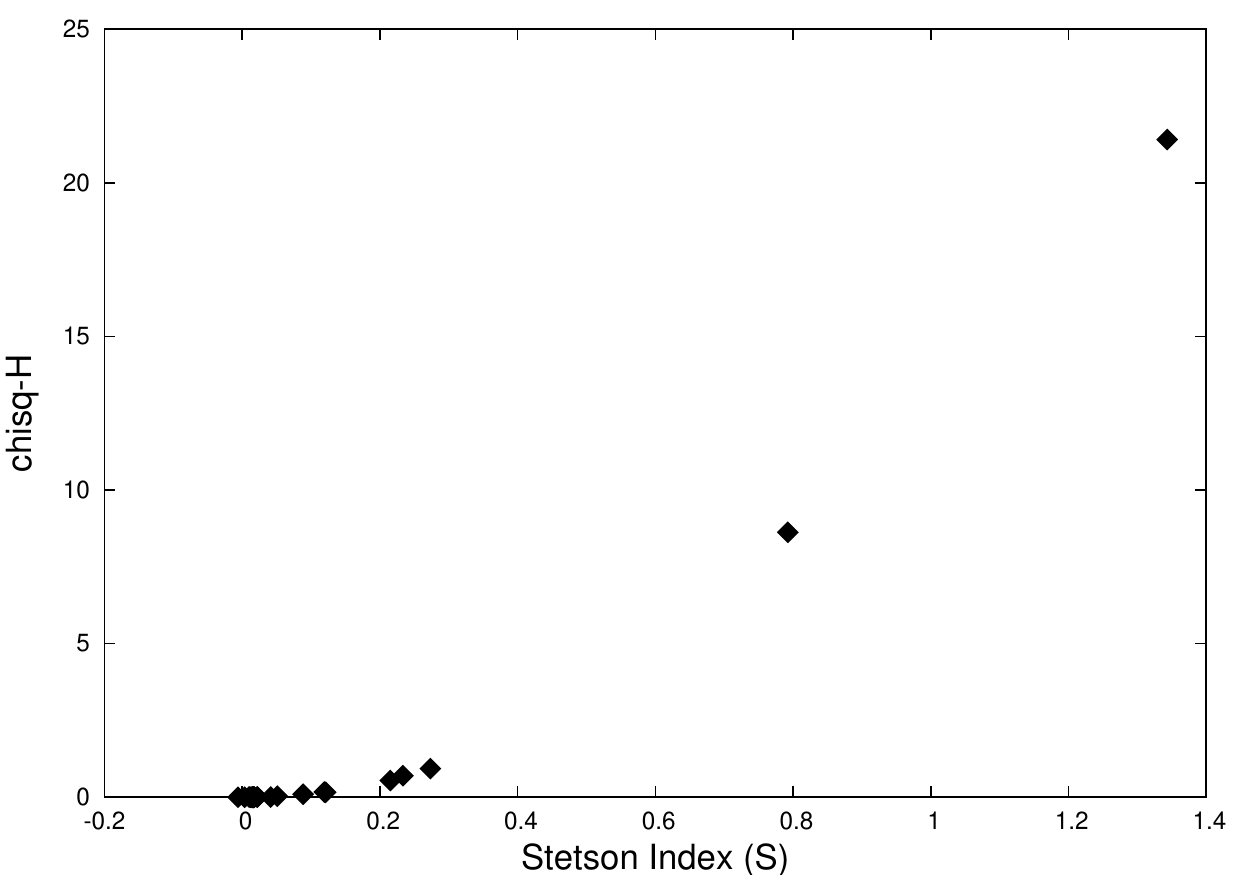} 
      \includegraphics[width=80mm]{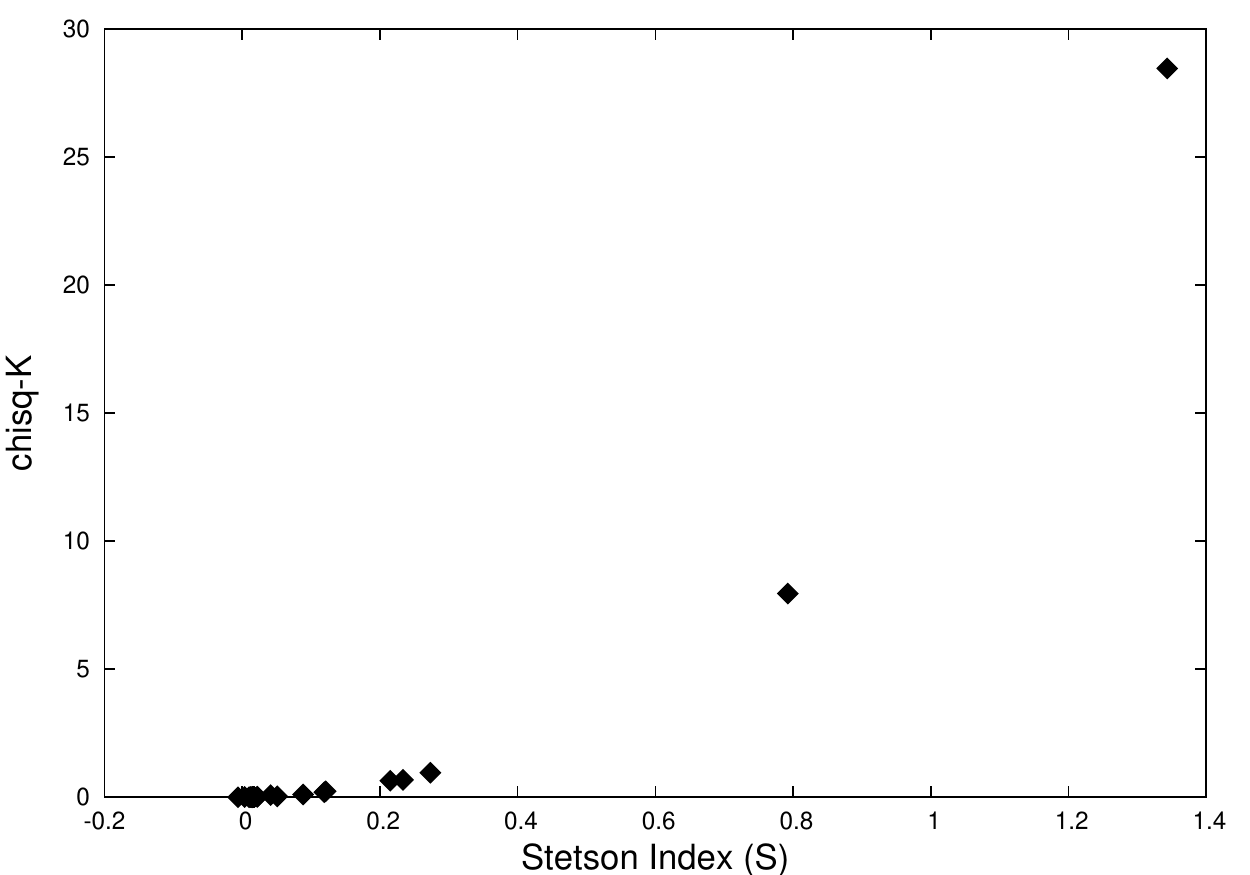} \\      
     \caption{{\it Top}: The Stetson variability index plotted against the $H$-band (2011) magnitude. {\it Bottom}: The reduced $\chi^{2}$ value of the $H$ (left) and $K_{s}$ (right) photometry for the reference sample, plotted against the Stetson index.  } 
    \label{stetson}
 \end{figure*}

Fig.~\ref{stetson} shows the index {\it S} plotted against the mean $H$-band magnitude for the reference stars, and C, E, and F. There is a trend of higher index for the brighter sources, as also noted by Carpenter et al. (2001). This can be expected, since the calculation of {\it S} includes dividing the differential magnitudes by the photometry error, and the brighter sources have smaller errors. The variability cut-off in the Stetson index can be taken from the reduced $\chi^{2}$ value of the observed magnitudes in each band, which was calculated as

\begin{equation}
\chi^{2} = \frac{1}{n-1} ~ \sum_{i=1}^{n} ~  \frac{(m_{i}-\bar{m})^{2}}{\sigma_{i}^{2}}.
\end{equation} 

A high $\chi^{2}$ value could result due to variability, or random noise. Plotting the $\chi^{2}$ values versus the Stetson index (Fig.~\ref{stetson}), the variability cut-off can be considered as $\chi^{2}$ $\geq$1 and $S \geq$ 0.2. Carpenter et al. have considered the variability cut-off to be $\chi^{2}$ $\geq$ 2 and $S \geq$ 0.55, but their sample also has a wider range in magnitudes and includes a high fraction of bright ($H <$ 14 mag) sources. We have used a lower index cut-off since we have a much fainter sample.

In Fig.~\ref{stetson}, there are just 5 sources that lie above the cut-off considered, and the two strongest contenders for variability (based on this criteria) are the two brightest sources (ID2 and ID15). As mentioned, there is a potential bias in the index being higher for brighter sources, which may not be variable. Also, these two objects lie close to the mean fit in the differential plots (Fig.~\ref{apfits}b), which would classify them as non-variable in both bands. For ID2, the peak counts are $\sim$55\% of the saturation limit in both bands, while for ID15, the peak counts are just at the level of $\sim$50\% of the saturation limit. The photometry for these two bright sources thus might be affected by non-linearity, which could result in a large Stetson index and high $\chi^{2}$ values. 

The Stetson variability index for ID11 is 0.12, below the threshold considered for variability (0.2), whereas it is much lower (0.02) for ID1. We note that the Stetson index is an indicator of correlated variability in multiple bands, and since ID 1 and 11 are both non-variable in the $K_{s}$-band, a low index value can be expected. Based on these arguments, we consider ID11 to be a long-term variable object, while ID1 might be marginally variable over the short-term period in the $H$-band. The comparatively higher standard deviation in the $H$-band differential magnitude found for ID1 could be due to a sudden fluctuation in this band, which is not correspondingly observed in the $K_{s}$ band. In the $K_{s}$-band differential magnitude plot, the object ID24 lies at a $\sim$1.87-$\sigma$ level above the mean rms (Fig.~\ref{apfits}b). This object is a non-variable in the $H$-band. We do not have the 2002 photometry for this source, since it lies out of the field of view of the CFHTIR observations. Therefore, we cannot check for any long-term variability. The Stetson index for ID24 is 0.015, which would classify it as a non-variable. Among the other sources in Fig.~\ref{stetson} which lie above index $S \geq$ 0.2, ID8, 9 and 12, are all consistent with being non-variables from the differential photometry plots, and also do not show any signs of long-term variability (Fig.~\ref{colors}). To summarize, we do not find any prominent short-term variable sources in our present dataset.

\subsection{Long-term NIR variability between 2002 and 2011} 
\label{longvar}

\subsubsection{Contrasting Blue/Red ($H-K_{s}$) Colors for TMR-1C}
\label{C} 

Table~\ref{photC} lists the photometry for TMR-1C from all previous and current observations. We find clear signatures of long-term variability when comparing the 2011 photometry with the one from 2002 and earlier epochs (Fig.~\ref{colors}). The amplitude of variations is different in the two bands. TMR-1C became brighter in the $H$-band by $\sim$1.7 mag between 1998 and 2002, and then fainter again by a smaller amplitude of $\sim$0.7 mag between 2002 and 2011. The $H$-band photometry is nearly constant between 1998 and 2000. In contrast, TMR-1C has persistently become brighter in the $K_{s}$-band in the $\sim$13 year period between 1998 and 2011. The brightening is of $\sim$0.4 mag between 1998 and 2002, and of a similar amplitude of $\sim$0.5 mag between 2002 and 2009. It is again found to be slightly brighter by $\sim$0.1 mag in the $K_{s}$-band in 2011. Overall, TMR-1C has become brighter by $\sim$1 mag in both bands over the 1998-2011 time period. 

%The amplitude of variation is larger when comparing with the {\it HST} photometry in both bands. 

The ($H-K_{s}$) color for TMR-1C shows large variations, from a nearly constant red value of 1.3--1.6 mag between 1998 and 2000, to a much bluer color of -0.1 mag in 2002, and then again a red color of 1.1 mag in 2011, similar to that observed in the 1998--2000 period. The variations are at a $>$2-$\sigma$ level. The period of variability might be $\sim$2 years, as that observed between 2000 and 2002, or shorter. The variations may also be aperiodic. As the ($H-K_{s}$) color gets bluer from 2000 to 2002, TMR-1C gets significantly brighter in the $H$-band (Table~\ref{photC}). When the color gets redder again between 2002 and 2011, TMR-1C gets fainter in $H$ but brighter in the $K_{s}$-band. The difference in the trends observed in the $H$- and $K_{s}$-band variability suggests the presence of more than one origin for the observed variations (Section \S\ref{origins}). 

%A greater number of measurements are required to constrain the variability period. 

%This brightening in the $K_{s}$-band compared to the trend in the $H$-band suggests the presence of more than one origin for the observed variations (Section \S\ref{origins}).

%h02vsh12
%k02vsk12

\subsubsection{Photometric Variability for TMR-1AB}

%There is no short-term variability observed for TMR-1AB over the day/week/month long periods probed in the 2011 observations. 
A comparison of the $H$-band photometry for TMR-1AB  indicates brightening by $\sim$2 mag between 1998 and 2002, and then the emission is nearly constant between 2002 and 2011 (Table~\ref{photC}). Based on radiative transfer modeling of the optical to millimeter SED for TMR-1AB, Furlan et al. (2008) concluded that the protobinary is viewed at a close-to edge-on inclination of $\sim$60--70$\degr$. The $H$-band variability observed between the 1998 and 2002 observations could be due to variable extinction by an edge-on disk or the ambient envelope material. It may also be the case that TMR-1AB is observed unocculted in all observations taken after 1998, which would explain the similar magnitudes in the 2002 and 2011 $H$-band observations. This protobinary is saturated in the $K_{s}$-band in the 2002 and 2011 observations, therefore the only ($H-K_{s}$) color measurement available is from the 2MASS observations in 1998. The {\it HST} colors from Terebey et al. (1998) are for the individual components of TMR-1AB, which makes it difficult to make a direct comparison with the 2MASS measurement. Within the uncertainties, the 1998 2MASS ($H-K_{s}$) color of 2.24 is similar to the {\it HST} color of $\sim$2-3 mag, although we would expect the composite photometry for the protobinary to be brighter than the individual components. Further observations are required to investigate if AB shows a similar blueing/reddening trend as observed for C. 

%For TMR-1AB, we have a ($H-K_{s}$) color of 1.1 mag from our 2002 photometry. From T98, the color is 2.3 mag for TMR-1A and 3.4 mag for TMR-1B. We would expect the composite photometry for the protobinary to be brighter, and so it is difficult to confirm any blueing observed in the 2002 observation for this system. This is a Class I system, surrounded by an infalling envelope and an actively accreting disk. Such systems are known to show strong veiling in the $K_{s}$-band, which would make the ($H-K_{s}$) color redder. Doppmann et al. (2005) estimate a mean $K_{s}$-band veiling for Class I systems in Taurus of $<r_{K}>$ = 1.54 $\pm$ 1.21. A variable mass infall rate could result in a variable ($H-K_{s}$) color. However, this system is also brighter in our 2002 observations, i.e. when the ($H-K_{s}$) colors are bluer. Disk emission then cannot be a possible origin of the variability. It is interesting to note that Chandler et al. (1999) have argued that TMR-1 is viewed at an intermediate inclination of $\sim$60$\degr$. Thus extinction by a (slightly) inclined disk or the ambient envelope material could be responsible for the variability, if indeed it is variable between the 1998 and 2002 observations. It seems more likely that we are observing TMR-1AB unocculted in all observations, if we return to the occulting scenario discussed in $\S$\ref{kh15d}, and the emission from the protobinary is more or less constant.

\subsubsection{Photometric variability for objects E and F}

The object F is a 2MASS source with a designation of 2MASS J04391199+2553490. A comparison of its 2MASS photometry with the 2002 and 2011 photometry indicates long-term variability in the ($H-K_{s}$) color of $\sim$1--1.4 mag (Table~\ref{photEF}). The object F gets bluer between 1998 and 2002 as it gets brighter in the $H$-band. This is a trend similar to TMR-1C, and is suggestive of the same physical origin for these sources. This is also accompanied by slight brightening of $\sim$0.1 mag in the $K_{s}$-band. It gets red again between 2002 and 2011 by a smaller amplitude of $\sim$0.4 mag as it gets fainter in the $H$-band. The emission in the $K_{s}$-band is nearly constant during this time period. In comparison, the object E shows smaller amplitude variation of $\sim$0.4 mag in the ($H-K_{s}$) color between 2002 and 2011, and as the color gets redder, E gets fainter in both bands by $\sim$0.1-0.5 mag (Table~\ref{photEF}). As noted in Section \S\ref{astrometry}, E is a strong contender for Taurus membership, whereas F appears as an outlier in the proper motion diagram shown in Fig.~\ref{pm}. There are WISE counterparts available for both E and F. However, the WISE mid-infrared images are strongly affected by the nebulosity surrounding TMR-1AB, and none of these sources could be resolved to obtain individual photometry. It is therefore difficult to confirm if E and F are disk sources, and if disk emission could be responsible for the observed variability.

\subsubsection{Two new variable sources}
\label{newvar}

\begin{figure*}    
 \centering
      \includegraphics[width=120mm]{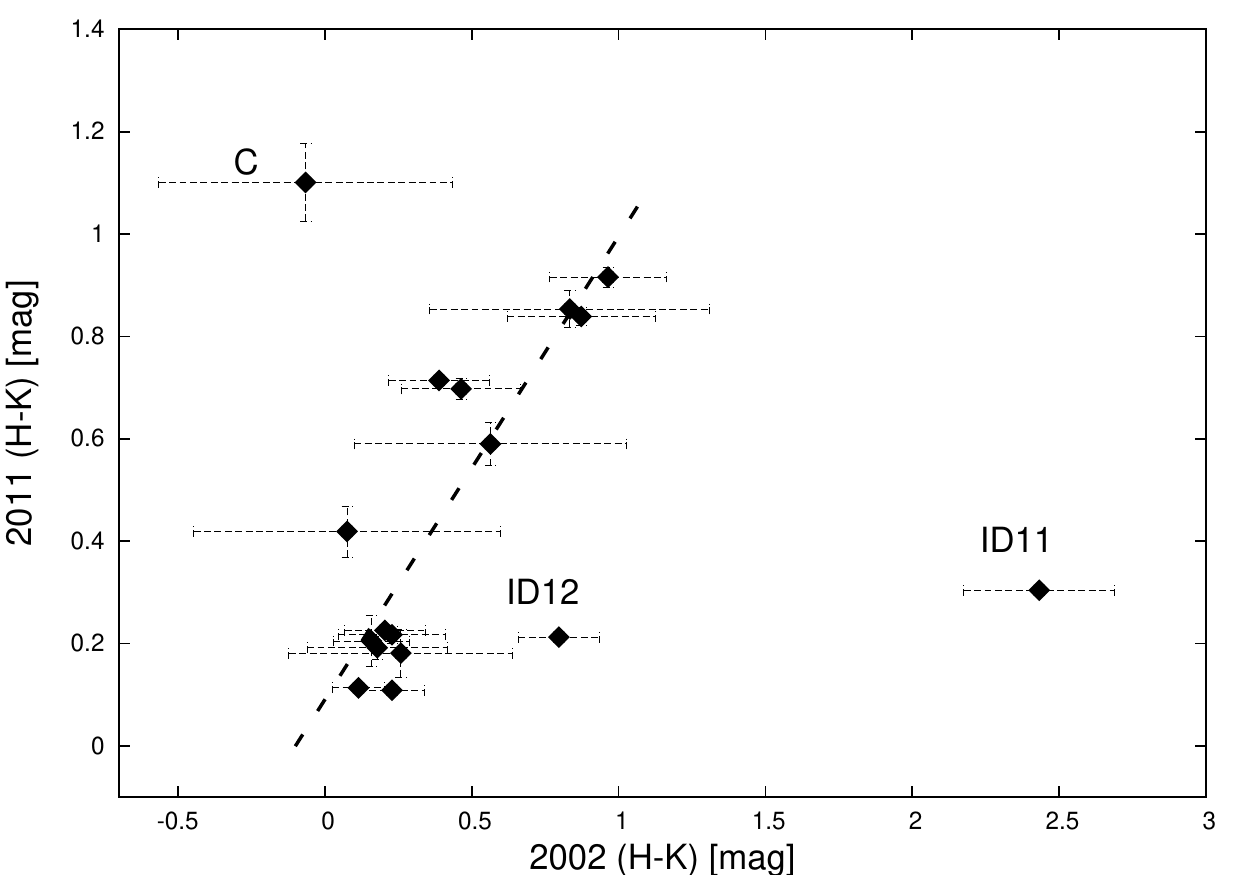}           
     \caption{ The 2011 versus the 2002 ($H-K_{s}$) color for the reference star sample and the targets. The dashed line is a linear fit to the reference star sample.  } 
    \label{colors}
 \end{figure*}

We discovered two new sources, ID11 and ID12, which show long-term variability when comparing the 2002 and 2011 ($H-K_{s}$) colors (Fig.~\ref{colors}). The photometry and astrometry for these objects are listed in Table~\ref{photEF} and Table~\ref{astroEF}, respectively. These are new detections and no matches were found in the SIMBAD database. Both objects are located in the south-west of TMR-1ABC (Fig.~\ref{spatial-full}). ID11, in particular, shows large-amplitude variation of $\sim$2 mag in the ($H-K_{s}$) color between 2002 and 2011, which is stronger than TMR-1AB or C. No short-term variability has been detected for this source in the 2011 data (Fig.~\ref{apfits}b). As the ($H-K_{s}$) color for this object gets bluer, the $H$ magnitude gets brighter by 1.7 mag, while the $K_{s}$ magnitude gets fainter by $\sim$0.3 mag (Table~\ref{photEF}). ID12, on the other hand, shows smaller amplitude variation of $\sim$0.5 mag between 2002 and 2011. There are no 2MASS matches or spectral classification available for these sources, however, we were able to find WISE counterparts for both objects. ID11 and ID12 are nice isolated sources in WISE 3.4 and 4.6 $\mu$m bands, with a detection of $>$10-$\sigma$, but are undetected in the 12 and 22 $\mu$m bands. These sources have a [3.4]-[4.6] color of 0.3--0.4 mag, and are likely Class III objects, although a few Class II sources with weak excesses in these shorter wavelength WISE bands exhibit bluer [3.4]-[4.6] colors than the majority of the disk sources (Riaz et al. 2012). Longer wavelength observations are required to confirm if ID11 and ID12 are disk bearing objects, and if the possible origin of variability could be disk emission or variable extinction. 

As can be seen in Fig.~\ref{pm}, the proper motion for ID11 is similar to TMR-1C, and is consistent with the Taurus loci, within the large error bars. ID12, however, appears as an outlier and may not be a candidate Taurus member. As discussed in Section \S\ref{astrometry}, with a longer baseline of 20 years or more, such outlier sources in the TMR-1 field might begin to separate from the Taurus loci, making it easier to determine their Taurus membership.

\section{Discussion}
\label{origins}

The absence of short timescale variability for TMR-1C suggests that it may not be a YSO. The typical variations in the NIR observed for protostars and T Tauri stars in Taurus are found to be $\sim$0.1--0.3 mag, with periods of a few days to a few weeks (e.g., Park \& Kenyon 2002). While we do not find any strong evidence of short-term variability upto a $\sim$4 month period for TMR-1C, based on the final accuracy of our present observations, we still cannot confirm small-amplitude variations of $<$0.15--0.2 mag in the NIR (Fig.~\ref{apfits}). The non-variability on a short timescale thus could be due to the amplitude level being lower than our detection limit. Several optical and NIR variability surveys of ultracool dwarfs in young clusters (e.g., Bailer-Jones et al. 2001; Carpenter et al. 2001; Joergens et al. 2003; Caballero et al. 2004; Scholz et al. 2004) have found amplitudes as small as 0.5--1\%. On the other hand, these surveys have noted that the fraction of variable sources is never higher than 30-40\%. The fraction of variable Class I stars in Taurus is also found to be similar (Park \& Kenyon 2002). Thus there can be YSOs which are not variable, and therefore the absence of short-term variability for TMR-1C, at any amplitude level, still does not reject this object as a candidate YSO. An important case to note is TMR-1AB itself, which does not show any short-term variability in the $H$-band, neither on a short- or long-term timescale (Fig.~\ref{apfits}; Table~\ref{photC}). It may also be the case that the variability period for C is longer than $\sim$4 months. The variable YSOs in Taurus and other young clusters have not been monitored over several years, and so long-term variations over a year or longer periods are not known for those objects. The various epochs of observations for TMR-1C compiled in Table~\ref{photC} are randomly sampled measurements, and the period of variability could be anywhere from a few months to the 2--4 years suggested by these observations. There is also the possibility that TMR-1C is going through a quiescent phase, and this object might be a case similar to AA Tau, which shows variability on several time-scales, from very large amplitudes to quiescent periods where the flux is stable over more than a week period (e.g., Bouvier et al.  2003). While we still do not have ample information to firmly conclude on the nature of TMR-1C, the arguments above provide a strong case for this source to be considered a potential YSO. 

%It has also been discussed extensively in previous works that TMR-1C could be a background star passing behind some patchy foreground material, and the cause of the long-term variations could be foreground extinction. 
There is also the possibility that TMR-1C is a background star passing behind some patchy foreground material, and the cause of the long-term variations could be foreground extinction. We argue that {\it if} foreground extinction is the main cause of the observed variations, then this should result in simultaneous brightening or dimming in both bands, at amplitudes roughly following the extinction law (e.g., A$_{H}$/A$_{K_{s}}$ $\sim$ 1.56; Reike \& Lebofsky 1985). As noted in Section \S\ref{C}, while C shows both dimming and brightening in the $H$-band at different amplitudes, it has persistently become brighter in the $K_{s}$ band between 1998 and 2011 (Table~\ref{photC}). The absence of any particular correlation observed between brightness and color changes is actually consistent with some variable YSOs in Taurus and the young $\sigma$ Orionis cluster (e.g., Park \& Kenyon 2002; Scholz et al. 2009). We would also expect variable foreground extinction to occur on a larger spatial scale, rather than simply affecting the object C. Our reference star sample is located roughly within 4$\arcmin\times$4$\arcmin$ of TMR-1C, but there are only two other objects in this region which show signs of long-term variability (Fig.~\ref{colors}), which may or may not be due to variable foreground extinction. We also note that the spectral energy distribution (SED) for TMR-1C cannot be fit using a model of an extincted background star, as shown in Petr-Gotzens et al. (2010). Any type of background star would have been detected in their ISAAC long wavelength images.

%It has been noted for ultracool dwarfs that they are not variable in the NIR at $\sim$0.2 mag amplitude levels (e.g., Radigan et al. 2012). 

 \begin{figure*}    
 \centering
      \includegraphics[width=150mm]{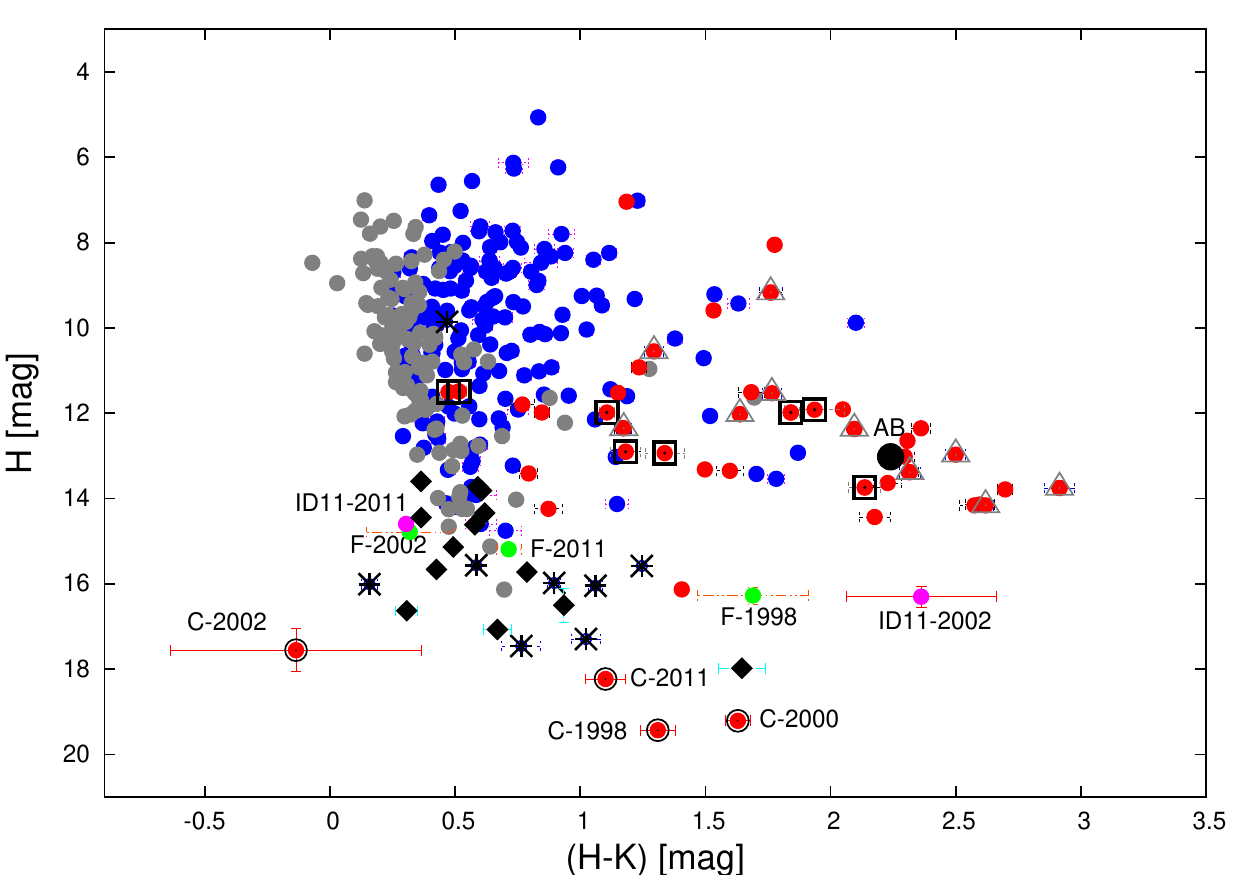}           
     \caption{A NIR color-magnitude diagram comparing the TMR-1 components and other variable sources with the Class I (red), Class II (blue), and Class III (grey) sources in Taurus. The protostellar sources with face-on ($<$50$\degr$) and edge-on ($>$50$\degr$) inclinations are marked by black open squares and grey triangles, respectively. Also plotted are 2011 colors for the two groups seen in Fig.~\ref{pm}, based on the proper motions; asterisks represent the group A with $\mu_{\alpha} >$0, filled diamonds represent the group B with $\mu_{\alpha} <$0.  } 
    \label{Taurus}
 \end{figure*}

The possible origins for the long-term variability observed for TMR-1C were discussed in detail in Riaz \& Mart\'{i}n (2011) and Petr-Gotzens et al. (2010), and we refer the readers to the discussion in these papers. What we have reconfirmed in our present observations is the large-amplitude long-term variations in the NIR emission, which were earlier observed between 1998 and 2002, and that no particular correlation is observed between the brightness and the color changes. This suggests that more than one origin is responsible for the observed variability for C. For e.g., the drop in magnitude in the $H$-band over the 2002-2011 period is $\sim$0.7 mag, which is much smaller than a $\Delta H$ of $\sim$1.7 mag observed during the 1998-2002 period. This brightening could be due to physical variations in the inner disk structure, such as a change in the thickness of the inner edge or certain local high energy events like accretion shocks that can cause thermal instabilities and raise the dust temperature in the inner disk region. Any structural inhomogeneities in a circumstellar disk, or even variable extinction due to the ambient molecular cloud, can cause the object to become bluer as it gets brighter in both bands, which is found to be the case for C during the 1998-2002 period. We again consider the example of AA Tau, which has been suspected to be surrounded by a warped disk undergoing dynamical structural changes and causing the variability timescales to change and/or to remain quiescent over different periods of time (e.g., Bouvier et al. 2003). There could also be variations in the magnetic field strength, which can modulate the accretion flow onto the inner disk. Such complicated inner disk geometry variations are expected to create aperiodic variations, with timescales varying from a few days to years (e.g., Carpenter et al. 2001).

There is, however, no concrete evidence yet that TMR-1C is surrounded by circumstellar material. This object was undetected in the IRAC mid-infrared bands, and in the ISAAC L' data (Petr-Gotzens et al. 2010). Therefore, we cannot measure its spectral slope so as to classify it. Given the faintness of this object, the disk would be too faint to be detected at the sensitivities of the IRAC and ISAAC instruments. It is important to note that Terebey et al. (2000) and Petr-Gotzens et al. (2010) have presented low-resolution NIR spectroscopic observations of TMR-1C. Both studies have found the NIR spectrum to be featureless. The spectra do not show any prominent absorption bands due to methane or water vapor, that dominate the spectra for cool objects at T$_{eff}$$<$ 2000 K. If TMR-1C is surrounded by a circumstellar disk which is at a high inclination angle to the line of sight, then the central object will be occulted by the disk, and its NIR spectrum will be dominated by the scattered light from the surface layers of the disk. The spectrum in such a case can be expected to appear featureless. While the mid-infrared data points for TMR-1C are all upper limits, radiative transfer modeling of the near- to mid-infrared SED suggests the presence of a highly inclined disk, at an inclination angle of 87$\degr$ (Petr-Gotzens et al. 2010). There are several degeneracies in the model fit, but the results are consistent with the featureless NIR spectrum. 

In Fig.~\ref{Taurus}, we have compared the NIR colors for TMR-1AB and C with other protostars and disk sources in the Taurus region. The Taurus Class I/II/III sources shown in this figure have been obtained from the work of Luhman et al. (2010). The ($H-K_{s}$) color obtained from the high-precision 1998, 2000 and 2011 photometry for TMR-1C are similar to the colors observed for the Taurus protostars, which are comparatively much brighter sources and have ($H-K_{s}$) $\geq$ 1. TMR-1C thus could be a very faint Class I source. Among these three measurements, there is variability in the ($H-K_{s}$) color of $\sim$0.3--0.5 mag, which is similar in amplitude to the variable YSOs in Taurus. The main discrepant for TMR-1C is the 2002 point, which is even bluer than the Class III sequence (Fig.~\ref{Taurus}). The typical intrinsic ($H-K_{s}$) color for Class I sources is estimated to be 0.6$\pm$0.4 (e.g., Doppmann et al. 2005). As discussed in Riaz \& Mart\'{i}n (2011), it may be that we have measured the intrinsic colors for TMR-1C in our 2002 measurement, while the 1998, 2000, and 2011 photometry is more affected by extinction. If indeed this blue color measurement is the intrinsic color for TMR-1C, then it could even be an ultracool brown dwarf, or a T dwarf, which is at an early Class 0/I evolutionary stage. T dwarfs are known to exhibit bluer colors than late-M or L dwarfs (e.g., Marley et al. 2002). We also note that the very low bolometric luminosity of  10$^{-3}$$L_{\sun}$ (T98) and high line of sight extinction of $A_{V}$$\sim$18 mag (Petr-Gotzens et al. 2010) are suggestive of this object being a very low-luminosity object (VeLLO; e.g., Dunham et al. 2008), although a correct classification of a VeLLO requires an estimate on the internal luminosity of the source, without any contribution from the surrounding envelope. To summarize, TMR-1C is a strong candidate for being a YSO, the classification of which requires further investigation.

% {\bf Given the much higher precision of the 1998, 2000, and the 2011 photometry compared to the 2002 measurement, we are more inclined to consider these redder colors. To summarize, TMR-1C should still be considered as a candidate YSO, the classification of which requires further investigation. }

%The 2002 color may represent the intrinsic color of this object, while the 1998, 2000, and 2011 observations may have been made while it passed behind some foreground material. On the other hand, considering the large error bar on the 2002 color, this point could be redder and lie at ($H-K_{s}$)$\sim$0.5, which would be more consistent with the other three measurements. 

The NIR color for TMR-1AB in Fig.~\ref{Taurus} is consistent with other protostellar sources in Taurus. As mentioned, TMR-1AB possesses a close to edge-on disk. For some of the known protostars in Taurus, the inclination angles have been estimated via radiative transfer modeling (e.g., Furlan et al. 2008). We have marked separately in Fig.~\ref{Taurus} the protostars that are viewed at face-on inclinations ($<$50$\degr$; open triangle) and edge-on inclinations ($>$50$\degr$; black open squares). There is no particular reddening observed in the NIR color for the edge-on sources; in fact, some of the face-on disks show redder colors than edge-on sources. This could be due to the difference in other disk/envelope parameters, such as a large cavity angle which can enhance NIR scattered emission even at intermediate inclinations, or due to different accretion properties (e.g., Riaz et al. 2009). Also plotted in Fig.~\ref{Taurus} are the 2011 colors for the two groups of stars that were noted in Section \S\ref{astrometry} to have different sets of proper motions (Fig.~\ref{pm}). It appears that the group A with $\mu_{\alpha} >$0 and which includes the object F, has slightly redder colors than group B, which includes the object E and mainly follows the main-sequence or the Class III locus. The objects in group A could be low-luminosity objects without a dusty disk excess emission, while the redder group B could consist of all disk sources. The difference in the colors of the two groups is very subtle, and requires further investigation. We note that the extension of the Class III sequence towards fainter magnitudes, as mentioned above, is more highlighted if we also consider the group A and group B stars plotted in Fig.~\ref{Taurus}, although some of these faint objects could be background stars and not all are expected to be YSOs. The objects F and ID11 show similar trends similar to C, with fainter magnitude in $H$-band as the ($H-K_{s}$) color becomes redder. Both of the 2002 and 2011 color measurements for the object E are consistent with the Class III locus (Table~\ref{photEF}). This could be either a photospheric or a transition disk source. The redder ($H-K_{s}$) color of $\sim$0.7 mag for ID12 suggests that this could be a Class II object. The WISE [3.4]-[4.6] colors for these sources are photospheric. Longer wavelength observations can correctly confirm if these could be transition disks with inner holes resulting in photospheric emission at near- and mid-infrared wavelengths.

\section{Summary}

We have conducted a near-infrared photometric monitoring of the TMR-1 system, and other objects in its vicinity. Our campaign was conducted with the CFHT/WIRCam imager in the $H$ and $K_{s}$ bands, between October, 2011, and January, 2012. We do not find TMR-1AB or TMR-1C to be variable in the NIR at amplitudes of $>$0.15--0.2 mag over the short-term period of $\sim$14 minutes to $\sim$4 months. We do find clear evidence of long-term variability for C, when comparing the present epoch of photometry with that available from 1998, 2000, 2002, and 2009. The amplitudes of variations are different for C in the $H$ and $K_{s}$ bands, which suggests that more than one mechanism could be the cause of the observed variability.

To summarize the nature of this object: (a) The non-detection of short-term variability is not inconsistent with TMR-1C being a YSO, considering that a large fraction of YSOs are found to be non-variable in the NIR. It may also be that the short-term variability is at an amplitude level below our detection limit ($\sim$0.2 mag), which would also be consistent with C being a YSO. (b) From the 1998, 2000, and 2011 photometry, which are at a much higher precision than the 2002 photometry, the ($H-K_{s}$) colors for C are similar to the protostars in Taurus, suggesting that it could be a faint dusty Class I source. (c) The absence of long-term variability on a large spatial scale, as well as the absence of simultaneous or correlated brightening/dimming in both $H$ and $K_{s}$ bands, argue against TMR-1C being a background star affected by foreground extinction. 

We have discovered a new object ($RA$=04:39:13.67; $Dec$=+25:53:47.47) with a proper motion of +0.66,-19.66 mas/yr (uncertainty of $\sim$30 mas/yr), similar to the typical proper motion for Taurus (+2,-22 mas/yr), making it a strong candidate for Taurus membership. Based on its faint magnitudes ($H$=17.6$\pm$0.02; $K_{s}$=17.2$\pm$0.04), this source is likely to be a brown dwarf. Our study has also revealed two new variable sources in the vicinity of TMR-1AB, which show long-term variations of $\sim$1--2 mag in the NIR colors between 2002 and 2011. The proper motions measured for TMR-1AB and TMR-1C are -40,+58 mas/yr and -22,+5 mas/yr, respectively, with an uncertainty of $\sim$30 mas/yr. A larger baseline of 20 years or more is required to confidently confirm the physical association of TMR-1AB and C.

\begin{acknowledgements}

EM was funded by Spanish ministry project AYA 2011-30147-C03-03. BR would like to thank visiting summer student C. Niven for his help with the data reduction. Based on observations obtained at the Canada-France-Hawaii Telescope (CFHT), which is operated by the National Research Council of Canada, the Institut National des Sciences de l'Univers of the Centre National de la Recherche Scientifique of France, and the University of Hawaii. Based on observations obtained with WIRCam, a joint project of CFHT, Taiwan, Korea, Canada, France, at the Canada-France-Hawaii Telescope (CFHT) which is operated by the National Research Council (NRC) of Canada, the Institute National des Sciences de l'Univers of the Centre National de la Recherche Scientifique of France, and the University of Hawaii. 

\end{acknowledgements}

\onecolumn

\begin{table}
\centering
\begin{minipage}{20cm}
\caption{Photometry for TMR-1AB and TMR-1C}
\begin{tabular}{ccccccccc}
\hline\hline

 TMR-1       & Observation (Epoch)   & Band/Filter  & Photometry & ($H-K_{s}$)\footnote{Using the color transformation relation from Persson et al. (1998); Eqn. (1)}   & $\Delta H$ & $\Delta K_{s}$ \\
 component &                          &            &  [mag]  & [mag]  &  [mag]  & [mag]    \\
\hline
C& WIRCam (Oct 2011 - Jan 2012)    &  $H$  &  18.23$\pm$0.03& 1.10$\pm$0.08 & 0.68$\pm$0.5\footnote{The difference of the 2011 and 2002 $H$- and $K_{s}$-band photometry. }  &  -0.55$\pm$0.5$^{b}$  \\
 & WIRCam (Oct 2011 - Jan 2012)   &  $K_{s}$ & 17.13$\pm$0.07 & && \\ \hline
C& WIRCam (Jan 2009)   &  $K_{s}$ & 17.2$\pm$0.3 &&& \\ \hline       
C& CFHTIR (Oct 2002)    &  $H$  &  17.55$\pm$0.5& -0.135$\pm$0.5 & -1.68$\pm$0.5\footnote{The difference of the 2002 and 1998 $H$- and $K_{s}$-band photometry. } & -0.44$\pm$0.5$^{c}$  \\
 & CFHTIR (Oct 2002)   &  $K_{s}$ & 17.68$\pm$0.5 & && \\ \hline
C& ISAAC (Oct 2000)\footnote{From Petr-Gotzens et al. (2010).}  & $H$ & 19.21$\pm$0.05 &1.63$\pm$0.05 &-0.22$\pm$0.05\footnote{The difference of the 2000 and 1998 $H$- and $K_{s}$-band photometry. }  & -0.6$\pm$0.07$^{e}$ \\
 & ISAAC (Oct 2000) & $K_{s}$ & 17.53$\pm$0.05 &  && \\ \hline
C& ISAAC (Dec 1998)$^{d}$ & $K_{s}$ & 17.63$\pm$0.1 &&& \\ \hline
C& {\it HST}/NICMOS\footnote{{\it HST}/NICMOS photometry from T98.} (Aug 1998)  & F160W (1.6$\mu$m) & 19.43$^{+0.04}_{-0.05}$ &1.31$\pm$0.07& &&  \\
 &  {\it HST}/NICMOS (Aug 1998) & F205W (2.05$\mu$m) & 18.12$\pm$0.06 & && \\   \hline
                                  
\hline

AB & WIRCam (Oct 2011 - Jan 2012)   &  $H$  & 11.04$\pm$0.02 & & -0.01$\pm$0.3$^{b}$ &   \\  \hline
AB & CFHTIR (Oct 2002)   &  $H$  & 11.05$\pm$0.3 &   &  -1.97$\pm$0.3$^{c}$  &  \\  \hline
A & {\it HST}/NICMOS$^{f}$ (Aug 1998) & F160W (1.6$\mu$m) & 15.65$^{+0.8}_{-0.5}$ &2.22$\pm$0.7 & & & \\
                                 &  {\it HST}/NICMOS (Aug 1998) & F205W (2.05$\mu$m) & 13.43$^{+0.4}_{-0.3}$ & & & \\ \hline
B & {\it HST}/NICMOS$^{f}$ (Aug 1998)   & F160W (1.6$\mu$m) & 17.22$\pm$1.1&3.44$\pm$1.0 & & & \\
                                 &  {\it HST}/NICMOS (Aug 1998) & F205W (2.05$\mu$m) & 13.78$^{+1.1}_{-0.4}$& & &  \\  \hline
AB & 2MASS (1998) & $H$ & 13.02$\pm$0.03 & 2.24$\pm$0.03 & & \\
       &  2MASS (1998)  & $K_{s}$ & 10.72$\pm$0.02 & & & \\  \hline

\hline

\end{tabular}
\label{photC}
\end{minipage}
\end{table}

\begin{table}
\centering
\begin{minipage}{20cm}
\caption{Astrometry for TMR-1AB and TMR-1C}
\begin{tabular}{ccccccccc}
\hline\hline
 TMR-1       & Observation (Epoch) & \multicolumn{2}{c}{Position}    & Sep\footnote{Separation[$\arcsec$] between TMR-1AB and TMR-1C. } & Position Error &  $\mu_{\alpha}$\footnote{Proper motion obtained from the 2002 and 2011 $H$-band positions. Errors on $\mu_{\alpha}$ and $\mu_{\delta}$ are 30.97 mas/yr and 30.90 mas/yr, respectively.  } & $\mu_{\delta}$  \\
component &                        &     RA (J2000) & Dec (J2000) &                      [$\arcsec$] & [$\arcsec$] & [mas/yr] & [mas/yr] \\
\hline

C   &  WIRCam (Oct 2011 - Jan 2012)   & 04 39 14.33 & +25 53 12.58   & 10.2   & 0.3-0.8 & -22.243  & +4.887     \\
AB   &  WIRCam (Oct 2011 - Jan 2012)   & 04 39 13.98 & +25 53 22.12   &    & 0.3-0.8 &  -40.027 & +57.95  \\ \hline

C     &  WIRCam (Jan 2009)   & 04 39 14.22 & +25 53 11.33   &    & 0.3-0.8 & & \\ \hline

C    & CFHTIR (Oct 2002) & 04 39 14.22 & +25 53 12.18    & 9.8 & 0.1 && \\ 
AB & CFHTIR (Oct 2002)  & 04 39 13.88 & +25 53 21.20  & & 0.1  & &   \\ \hline

C & {\it HST}/NICMOS\footnote{{\it HST}/NICMOS positions from T98.} (Aug 1998)  & 04 39 14.14 & +25 53 11.8 &  10 & 0.35 & &  \\   
A & {\it HST}/NICMOS (Aug 1998)  & 04 39 13.84 & +25 53 20.6 & & 0.35 \\ 
B & {\it HST}/NICMOS (Aug 1998) & 04 39 13.83 & +25 53 20.4  & & 0.35& \\  \hline
AB & 2MASS (1998) & 04 39 13.89 & +25 53 20.88 & & 0.08 & &  \\ 
                                          
\hline\hline
\end{tabular}
\label{astroABC}
\end{minipage}
\end{table}

\begin{table}
\centering
\begin{minipage}{20cm}
\caption{Photometry for Objects E, F, and variable candidates}
\begin{tabular}{ccccccccc}
\hline\hline
 Object       & Observation (Epoch)   & Band/Filter  & Photometry & ($H-K_{s}$)  \\
     &                             &            &  [mag]  & [mag]     \\
\hline
E    & WIRCam (Oct 2011 - Jan 2012)    &  $H$  &  17.632$\pm$0.023& 0.41 $\pm$0.05 &  \\
       &  WIRCam (Oct 2011 - Jan 2012)   &  $K_{s}$ & 17.2127$\pm$0.044 & \\ 
      
E    & CFHTIR (Oct 2002)    &  $H$  &  17.099$\pm$0.4& 0.006$\pm$0.5 &  \\
       &  CFHTIR (Oct 2002)   &  $K_{s}$ & 17.093$\pm$0.336 & \\ \hline

F    & WIRCam (Oct 2011 - Jan 2012)    &  $H$  &  15.188$\pm$0.003& 0.7144$\pm$0.005 &  \\
       &  WIRCam (Oct 2011 - Jan 2012)   &  $K_{s}$ & 14.4736$\pm$0.004 & \\ 
      
F    & CFHTIR (Oct 2002)    &  $H$  &  14.7824$\pm$0.14& 0.3189$\pm$0.17 &  \\
       &  CFHTIR (Oct 2002)   &  $K_{s}$ & 14.4635$\pm$0.098 & \\   
       
F    & 2MASS (1998)    &  $H$  &  16.27$\pm$0.20& 1.69$\pm$0.22 &  \\
       & 2MASS (1998)   &  $K_{s}$ & 14.58$\pm$0.1 & \\    \hline    
       
ID1    & WIRCam (Oct 2011 - Jan 2012)    &  $H$  &  15.7972$\pm$0.017& 0.6977$\pm$0.02 &  \\
       &  WIRCam (Oct 2011 - Jan 2012)   &  $K_{s}$ & 15.0995$\pm$0.013 & \\ 
      
ID1    & CFHTIR (Oct 2002)    &  $H$  &  15.3994$\pm$0.16& 0.3939$\pm$0.2 &  \\
       &  CFHTIR (Oct 2002)   &  $K_{s}$ & 15.0055$\pm$0.126 & \\  \hline
       
ID2    & WIRCam (Oct 2011 - Jan 2012)  &  $H$  &  16.8282$\pm$0.037& 0.1815$\pm$0.05 &  \\
       &  WIRCam (Oct 2011 - Jan 2012)   &  $K_{s}$ & 16.6467$\pm$0.03 & \\ 
      
ID2    & CFHTIR (Oct 2002)    &  $H$  &  16.6394$\pm$0.3& 0.1889$\pm$0.38 &  \\
       &  CFHTIR (Oct 2002)   &  $K_{s}$ & 16.4505$\pm$0.25 & \\         \hline          

ID11    & WIRCam (Oct 2011 - Jan 2012)    &  $H$  &  14.593$\pm$0.004& 0.3042$\pm$0.008 &  \\
       &  WIRCam (Oct 2011 - Jan 2012)   &  $K_{s}$ & 14.2889$\pm$0.007 & \\ 
      
ID11    & CFHTIR (Oct 2002)    &  $H$  &  16.3014$\pm$0.245& 2.3619$\pm$0.3 &  \\
       &  CFHTIR (Oct 2002)   &  $K_{s}$ & 13.9395$\pm$0.1 & \\ \hline

ID12    & WIRCam (Oct 2011 - Jan 2012)    &  $H$  &  14.0096$\pm$0.004& 0.2127$\pm$0.006 &  \\
       &  WIRCam (Oct 2011 - Jan 2012)   &  $K_{s}$ & 13.7969$\pm$0.005 & \\ 
      
ID12    & CFHTIR (Oct 2002)    &  $H$  &  14.6684$\pm$0.11& 0.7269$\pm$0.13 &  \\
       &  CFHTIR (Oct 2002)   &  $K_{s}$ & 13.9415$\pm$0.1 & \\    \hline
       
ID15    & WIRCam (Oct 2011 - Jan 2012)  &  $H$  &  12.4135$\pm$0.03& 0.253$\pm$0.04 &  \\
       &  WIRCam (Oct 2011 - Jan 2012)   &  $K_{s}$ & 12.1606$\pm$0.03 & \\  \hline   %2002 out of frame
       
ID24    & WIRCam (Oct 2011 - Jan 2012)  &  $H$  &  17.5247$\pm$0.03& 1.1322$\pm$0.05 &  \\
       &  WIRCam (Oct 2011 - Jan 2012)   &  $K_{s}$ & 16.3925$\pm$0.04 & \\    %2002 out of frame       

\hline\hline
\end{tabular}
\label{photEF}
\end{minipage}
\end{table}

\begin{table}
\centering
\begin{minipage}{20cm}
\caption{Astrometry for Objects E, F, ID11, and ID12}
\begin{tabular}{ccccccccc}
\hline\hline
 Object       & Observation (Epoch) & \multicolumn{2}{c}{Position}    &  $\mu_{\alpha}$\footnote{Proper motion obtained from the 2002 and 2011 $H$-band positions. Errors on $\mu_{\alpha}$ and $\mu_{\delta}$ are 30.97 mas/yr and 30.90 mas/yr, respectively.  } & $\mu_{\delta}$  \\
 &                        &     RA (J2000) & Dec (J2000) &  [mas/yr] & [mas/yr] \\
\hline

E   &  WIRCam (Oct 2011 - Jan 2012)   & 04 39 13.6735 & +25 53 47.476    & +0.6599   & -19.665      \\
E   & CFHTIR (Oct 2002)   & 04 39 13.5465 & +25 53 47.328     &   &      \\ \hline

F   &  WIRCam (Oct 2011 - Jan 2012)   & 04 39 12.1618 & +25 53 48.820    & +63.005  & -45.11699       \\
F   & CFHTIR (Oct 2002)   & 04 39 11.9933 & +25 53 48.928     &   &      \\ 
F   & 2MASS (1998)   & 04 39 11.990 & +25 53 49.06     &   &     \\ \hline

ID11   &  WIRCam (Oct 2011 - Jan 2012)   & 04 39 05.648 & +25 51 42.586      & -26.52699  & +15.18299    \\
ID11  & CFHTIR (Oct 2002)   & 04 39 05.5393 & +25 51 42.090    &   &    \\ \hline

ID12   &  WIRCam (Oct 2011 - Jan 2012)   & 04 39 05.7054 & +25 51 30.399  & +0.29299  & +76.995      \\
ID12   & CFHTIR (Oct 2002)   & 04 39 05.5787 & +25 51 29.285    &   &     \\  %\hline

%ID1   &  WIRCam (Oct 2011 - Jan 2012)   & 04 39 14.9944 & +25 54 17.851  & 30.029  & -29.349     \\
%ID1   & CFHTIR (Oct 2002)   & 04 39 14.8479 & +25 54 17.801   &   &      \\ 

\hline\hline
\end{tabular}
\label{astroEF}
\end{minipage}
\end{table}

\end{document}